\documentclass{aa501}
\usepackage{graphicx}
\begin{document}
   \title{An artificial meteors database as a test for the presence of 
weak showers}

   \author{Arkadiusz Olech
          \inst{1} and Mariusz Wi\'sniewski \inst{2}
          }

   \offprints{A.Olech}

   \institute{Nicolaus Copernicus Astronomical Center, ul. Bartycka 18,
	00-716 Warszawa, Poland \\
	\email{olech@camk.edu.pl}
	\and
	Warsaw University Observatory
              Al.Ujazdowskie 4, 00-478 Warszawa, Poland\\
              \email{mwisniew@sirius.astrouw.edu.pl}
             }

   \date{Received .................., 2001; accepted ................ 2002}

   \abstract{We have constructed an artificial meteor database
resembling in all details the real sample collected by the observers of
the {\sl Comets and Meteors Workshop} in the years 1996-1999. The artificial
database includes the sporadic meteors and also events from the following
showers: Perseids, Aquarid complex, $\alpha$-Capricornids, July Pegasids
and Sagittarids. This database was searched for the presence of the
radiants of two weak showers: $\alpha$-Cygnids and Delphinids. The lack
of these radiants in the artificial database and their existence in the real
observations suggests that $\alpha$-Cygnids and Delphinids are the real
showers and their radiants could not be formed as an effect of
intersections of back prolongated paths of meteors belonging to other
showers.
   }

\titlerunning{An artificial meteors database as a test ...}
   \maketitle
%

\section{Introduction}

Recently, Polish observers taking part in the {\sl Comets and Meteors
Workshop (CMW)} have reported the rediscovery of two July meteor showers -
$\alpha$-Cygnids and Delphinids (Olech et al. 1999a, 1999b, Stelmach \&
Olech 2000, Wi\'sniewski \& Olech 2000, 2001). Both of these showers are
weak with maximum Zenithal Hourly Rates (ZHRs) slightly exceeding or
laying beneath the sporadic background. 

The $\alpha$-Cygnids are active from the end of June until the end of
July. The highest activity with ${\rm ZHR}=2.4\pm0.1$ is observed at
a solar longitude $\lambda_\odot=114.8\pm0.5^\circ$. The radiant of the
shower at this moment is placed at $\alpha=305^\circ$ and
$\delta=+45^\circ$.

The activity period of the Delphinids is still quite uncertain with the
first meteors from this shower detected around July 10 and the last ones
as late as the middle of August. According to the recent work of
Wi\'sniewski \& Olech (2001) the maximum hourly rates are observed at
$\lambda_\odot=125.0\pm0.1^\circ$. The activity at this moment is equal
to ${\rm ZHR}=2.2\pm0.2$ and the radiant of the shower has the
equatorial coordinates equal to $\alpha=312^\circ$ and $\delta=+12^\circ$.

The equatorial coordinates of the beginnings and ends of meteor paths and
its angular velocities for both showers were carefully analyzed using
the {\sc radiant} software (Arlt 1992). This software takes into
account the properties of the observed meteors and computes the
maps of probability for the presence of a radiant (hereafter PPR maps).

Although PPR maps computed for both of these showers showed distinct
features, the resulting radiants were polluted by the influence of the
meteors from other showers. A quite strong tail reaching the radiant of
the Perseids was detected in the case of the $\alpha$-Cygnids. There is
also a trace of the weak $o$-Draconids radiant in the close vicinity of
the radiant of the $\alpha$-Cygnids (Olech et al, in preparation).

An even more complicated situation is present in the case of the
Delphinids. The radiant of this shower is placed not far from the series
of ecliptic radiants of the Aquarids complex, $\alpha$-Capricornids and
the Sagittarids.

The radiants of these showers are large and have a complex structure often
showing several maxima of activity.

Thus one can suspect that both the $\alpha$-Cygnids and Delphinids are not
the real showers and their radiants produced by {\sc radiant} software
come from crossing the back-prolongated paths of the meteors from other
showers active in July and also from sporadic events.

To check this possibility, we decided to construct a realistic database of
artificial meteors which thoroughly resembled the real sample analyzed in
the above mentioned papers.

\section{July showers}

There are many meteor showers active in July. The most active of them
are the Perseids and $\delta$-Aquarids S but there are also several minor
showers such as the Sagittarids, the July Pegasids, $\alpha$-Capricornids, 
$\delta$-Aquarids N and the $\iota$-Aquarids N (Rendtel et al. 1995).

Recent compilations of meteor shower activity were done by Jenniskens
(1994) and Rendtel et al. (1995). The first of these papers presented
the results obtained in the years 1981-1991 and the second the results from
the period 1988-1995. On the other hand, the databases used by Olech et al.
(1999a, 1999b), Stelmach \& Olech (2000) and Wi\'sniewski \& Olech
(2000, 2001) were obtained using observations collected in the years
1996-1999. Because this period does not overlap with those mentioned
above we decided to make a new compilation of meteor activity in July.

We used the Visual Meteor Databases (VMDB) constructed each year by the
{\sl International Meteor Organization (IMO)} and accessible via the
{\sl IMO} web pages (Arlt 1997, 1998, 1999, 2000).

The observations from the databases were selected according to the
following criteria:

\begin{itemize}

\item data with limiting magnitudes less than 5.5 are omitted,

\item observing intervals should be longer than 0.5 hour,

\item radiant altitude should be at least 20 degrees,

\item field of view has less than 50\% cloud obstruction,

\end{itemize} 

The numbers of meteors from each shower, the numbers of sporadic events,
The activity periods used for computation of activity profiles and the
effective time of observations in these periods are summarized in Table
1.

\begin{table}[h]
\caption{VMDB statistics for each July shower from years 1996-1999}
\begin{tabular}{|l|c|c|c|c|}
\hline
\hline
Shower & Activity period & $N_{\rm met}$ & $N_{\rm spor}$ & $T_{\rm eff}$ \\
\hline
\hline
Sagittarids & Jul 1-15 & 302 & 3072 & 400.6 \\
July Pegasids & Jul 5-15 & 649 & 5237 & 612.8 \\
$\alpha$-Capricor. & Jul 1 - Aug 25 & 5403 & 68735 & 7469.3 \\
$\delta$-Aquarids N & Jul 10 - Aug 28 & 4091 & 48782 & 5244.9 \\
$\delta$-Aquarids S & Jul 6 - Aug 22 & 3841 & 28355 & 2746.6 \\
$\iota$-Aquarids S & Jul 14 - Aug 25 & 1241 & 25808 & 2404.3 \\
Perseids & Jul 13 - Aug 5 & 13953 & 38060 & 3937.2 \\
\hline
\hline
\end{tabular}
\end{table}

Following the example of Jenniskens (1994) we
fitted the following equation format to each activity profile:

\begin{equation}
{\rm ZHR}={\rm ZHR}_{\rm max}\cdot
10^{-B|\lambda_\odot-\lambda^{max}_\odot|}
\end{equation}

\noindent where $\lambda_\odot$ denotes the solar longitude for the epoch of
2000.0, and Zenithal Hourly Rates (ZHRs) are computed as follows:

\begin{equation}
{\rm ZHR}={{N\cdot F\cdot r^{(6.5-LM)}}\over{T_{\rm eff}\cdot\sin(H_{\rm
rad})}}
\end{equation}

\noindent where $N$ is a number of meteors observed during $T_{\rm
eff}$, $F$ is the cloud correction factor, $H_{\rm rad}$ is the altitude
of the radiant, $LM$ is the limiting magnitude in the field of view and
$r$ is the population index, which values are taken from Rendtel et al.
(1995).

For each of the analyzed showers we computed ${\rm ZHR}-\lambda_\odot$
dependence and we fitted it using the formula given in equation (1).
Free parameters ${\rm ZHR}_{\rm max}$, $\lambda^{max}_\odot$ and $B$ were
determined using the least squares method.

\begin{table}[t]
\caption{${\rm ZHR}_{\rm max}$, $\lambda^{max}_\odot$, $B$ parameters
and assumed radiant radii $R$ for July showers}
\begin{center}
\begin{tabular}{|l|r|r|r|c|}
\hline
\hline
Shower & ${\rm ZHR}_{\rm max}$ & $\lambda^{max}_\odot$ [$^\circ$]& $B$
[$1/^\circ$] & $R$ [$^\circ$]\\
\hline
\hline
July Pegasids & 3.11 & 108.52 & 0.0760 & 0.8 \\
               & $\pm$0.13 & $\pm$0.24 & $\pm$0.0101 & \\
\hline
$\alpha$-Capricornids & 3.41 & 126.23 & 0.0352 & 2.5 \\
               & $\pm$0.05 & $\pm$0.17 & $\pm$0.0008 & \\
\hline
$\delta$-Aquarids N & 2.62 & 130.03 & 0.0213 & 1.0 \\
               & $\pm$0.05 & $\pm$0.33 & $\pm$0.0009 & \\
\hline
$\delta$-Aquarids S & 8.99 & 127.05 & 0.0666 & 1.0 \\
               & $\pm$0.16 & $\pm$0.11 & $\pm$0.0013 & \\
\hline
$\iota$-Aquarids S & 2.49 & 126.92 & 0.0491 & 1.0 \\
               & $\pm$0.08 & $\pm$0.27 & $\pm$0.0020 & \\
\hline
Sagittarids & 3.00 & 90.00 & 0.0351 & 3.0 \\
               & -- & -- & $\pm$0.0015 & \\
\hline
Perseids & 17.0 & 136.70 & 0.0430 & 0.8 \\
               & -- & -- & $\pm$0.0010 & \\       

\hline
\hline
\end{tabular}
\end{center}
\end{table}

The results obtained for all July showers are presented in Table 2 and
their activity profiles drawn using the equation (1) shown in Fig.
1.

Two showers from our sample required special treatment. As was shown by
Jenniskens (1994) in the case of the Perseids, a nearly exponential
increase in activity is observed only between $\lambda_\odot=120^\circ$
and $\lambda_\odot=137^\circ$ with $B=0.050\pm0.005$ and ZHR at the end
of this period is equal to 18. After $\lambda_\odot=137^\circ$ the slope
$B$ changes to $0.20\pm0.01$, and after $\lambda_\odot=141.8^\circ$
changes again to $0.083\pm0.017$. Thus, we cannot describe the activity
of the Perseids using only one formula in the form of (1). Fortunately
we are interested only in the Perseids activity in July and therefore we
analyzed data for this stream only for $\lambda_\odot<137^\circ$. For
this shower we obtained that at $\lambda_\odot=136.7^\circ$ the ${\rm
ZHR_{max}}$ is equal to 17.0. Thus in the case of the Perseids, the only
free parameter in the equation (1) is $B$.

Another unusual shower is the Sagittarids for which Rendtel et al.
(1995) found several maxima of activity. In this case, we computed the
ZHR for the beginning of July and assumed this moment as the maximum.
Thus again the only free parameter was the slope $B$.

A completely different approach we performed in the case of sporadic
meteors. As was shown in Znojil (1995) the hourly rates of sporadic
events increase almost linearly between 22 and 2 at local time with a
slope equal to 2.1 meteors per hour. Using VMDB of IMO we also computed
the mean value of HR for sporadic meteors in July. Thus

   \begin{figure}[h]
   \centering
\includegraphics[scale=.90]{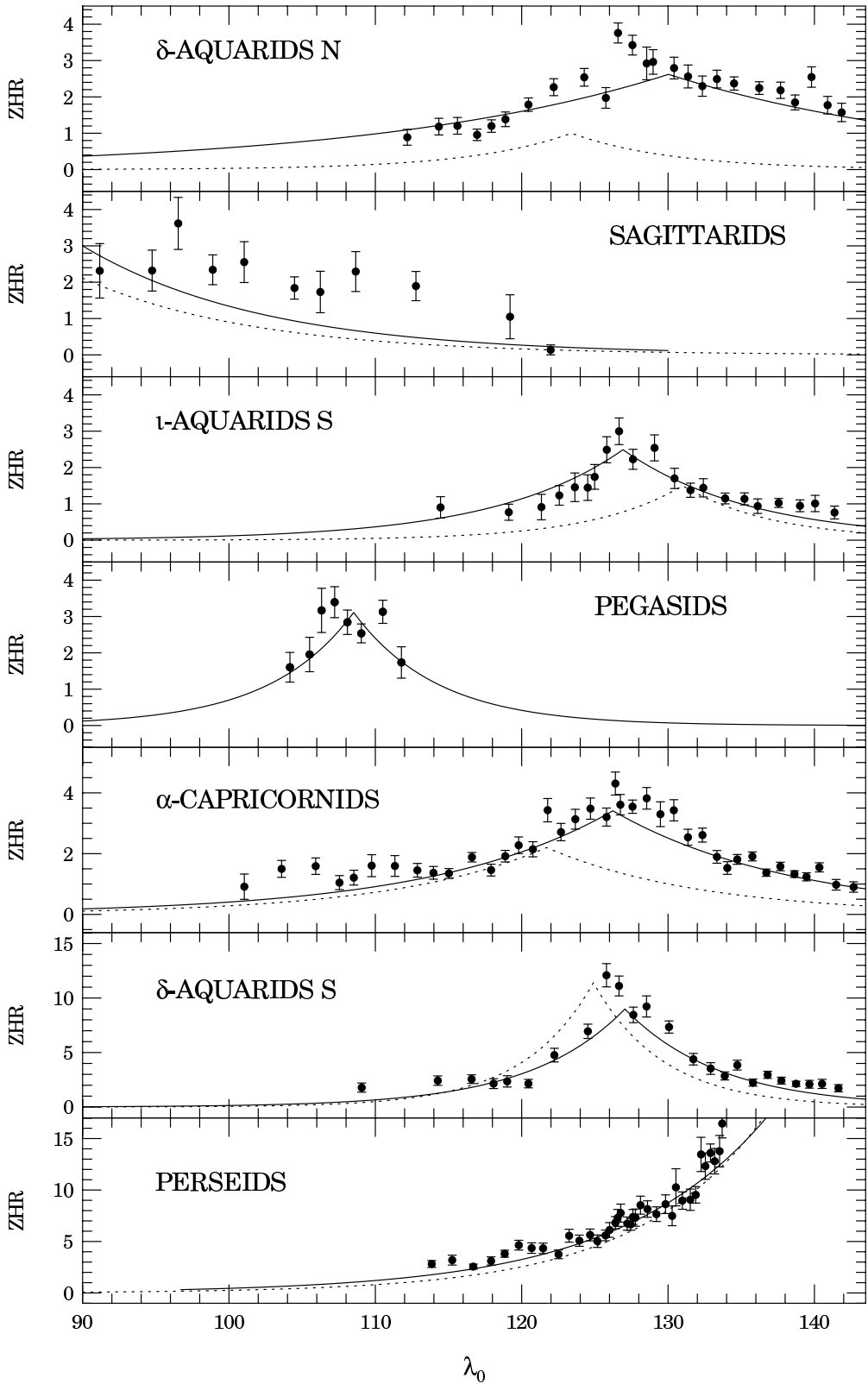}
      \caption{Activity profiles for July showers. The solid lines
correspond to the fits based on the equation (1). The dotted lines are
the activity profiles given by Jenniskens (1994)
              }
         \label{FigVibStab}
   \end{figure}

\clearpage

\noindent the sporadic background for each night was described by the
following equation:

\begin{equation}
{\rm HR}_{spor}=2.12\cdot UT-33.64
\end{equation}

\section{Observations}

The sample analyzed by Wi\'sniewski \& Olech (2001) contained 1372 hours of
effective time of observations collected in 1996-1999. Due to the poor
weather conditions in Poland these observations were not distributed
uniformly. Thus the artificial database, which we want to construct should
take into account the distribution of real observations. This distribution
is presented in Fig. 2.

   \begin{figure}[h]
   \centering
\includegraphics[scale=.75]{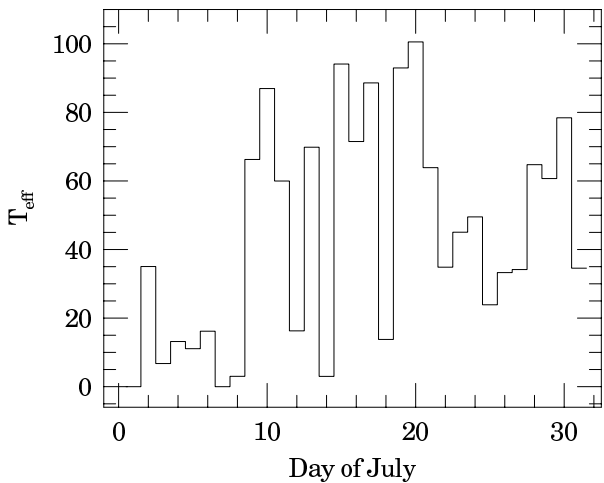}
      \caption{Distribution of real observations collected by the
{\sl CMW} observers in the years 1996-1999
              }
         \label{FigVibStab}
   \end{figure}

Due to the fact that the majority of the {\sl CMW} observations were made
during the astronomical camps, which took place in the Observational
Station of Warsaw University Observatory in Ostrowik, we assumed that
all observations in the artificial sample were collected in Ostrowik
($\lambda=21.4^\circ$~E, $\phi=52.1^\circ$~N). We also assumed that the
limiting magnitude $LM$ was the same for all observations and was at the
level of 6.3 mag., a value typical for Ostrowik conditions.

Knowing the number of observations collected on each July night we
distributed them uniformly from 20:00 to 24:00 UT - the period of time
in which the Sun is sufficiently enough below the horizon for meteor
observations in Poland.

For each moment, using values from Table 2 and the equation (1), we can
compute the expected value of ZHR for each shower. Knowing the ZHRs, and
also the values of altitudes of the radiants for each hour, and using
equation (2), we can calculate the expected numbers of observed meteors
$N$. Of course, due to the different perception of the observers, some
of them, even in the same conditions, detect a higher number of meteors
than others. Thus we modified the expected numbers $N$ using Gaussian
distribution with the mean at $N$ and the standard deviation equal to
$\sqrt N$.

As a result we obtained a file containing the date of the observation,
the UT time at the middle of the observation and the numbers of meteors from
each shower and also the number of sporadic meteors.

Finally in our artificial sample we included 1042 $\alpha$-Capricornids,
750 $\delta$-Aquarids N, 605 July Pegasids, 803 $\delta$-Aquarids S, 396
$\iota$-Aquarids S, 121 Sagittarids, 2465 Perseids and 15394 sporadics.
This sample is clearly more numerous than the real database. It is due
to the fact that during the construction of the artificial sample we
assumed a constant and quite high value of the limiting magnitude
($LM=6.3$ mag.) and additionaly assumed that all artificial observations
were taken under a clear sky.

\section{Distribution of shower meteors over the celestial sphere}

\subsection{Locations of meteor paths}

Due to the small perturbations caused by the bodies of the Solar System,
paths of the meteoroids from a particular stream are not ideally
parallel in the Earth atmosphere. Thus the radiants of meteor showers
are not the ideal points. According to the recent video results (Molau
2000, 2001) the radiant radii of the Perseids and the Leonids are around
1 degree. For more complex ecliptic radiants like the
$\alpha$-Capricornids, the Sagittarids and the Taurids, these radii are
around 3 degrees (Molau 2000, Triglav 2001). In our artificial database
we have to take into account these radiant sizes. Thus for showers like
the Perseids and the July Pegasids, which are rich in young material,
assumed radiant radii were equal to 0.8 degree. Other normal showers
have these radii equal to 1 degree. For the Sagittarids and the 
$\alpha$-Capricornids we assumed radii equal to 3.0 and 2.5 degrees,
respectively. These values are summarized in Table 2.

   \begin{figure}[h]
   \centering
\includegraphics[scale=.5]{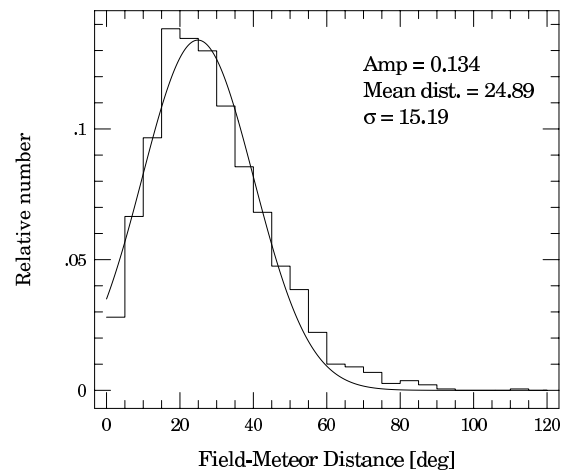}
      \caption{The distribution of distances between the center of the  
observed field and the beginning of the meteor path for real data
collected by the {\sl CMW} observers
              }
         \label{FigVibStab}
   \end{figure}


Knowing the theoretical radiant center and its radius, for each event
from a given shower we calculated its real radiant position using the two
dimensional Gaussian surfaces with a center at the theoretical radiant
center and standard deviations in right ascension and declination equal
to the radiant radius given in Table 2.

During the real observation each observer of the {\sl CMW} was looking
in a specific direction. These details are noted in observational
report form by giving the equatorial coordinates of the center of the field
of view. This center should be always at an elevation of at least 40
degrees.

The artificial database should take into account a location of the field
of view of an observer. Therefore we constructed a list of the centers
of view used in the real sample. In the first step, for each artificial
observation, we randomly drew a center of the field of
view from our list.

Having a center we can analyze the distribution of the meteors in the
field of view. Using the real observations we calculated the distances
of the beginnings of the meteor paths from the center of the field of
view. The distribution of these distances is presented in Fig. 3. The
best fit to this distribution was obtained using the Gaussian function
with the mean value equal to $24.9^\circ$ and $\sigma=15.2^\circ$, shown
as a solid line in Fig. 3.

Of course meteors appear irrespective of the distance from the
field of view. Another factor which we should take into account is the
location of the radiant. Thus, finding the beginning of the meteor path
was done as follows:

\begin{enumerate}

\item knowing the center of the field of view, the distance of the
beginning of the meteor path was drawn using the distribution presented
in Fig. 3,

\item the line between the center of the field of view and the radiant
of the shower was found,

\item the angle between the line mentioned above and the line connecting
the center of the field of view with the beginning of the meteor path was
found using the Gaussian distribution with a mean equal to zero and
a standard deviation equal to $60^\circ$,

\item to have a quite uniform distribution of meteors around the center of
the field in 40\% of the cases, the angle mentioned above was increased by
$180^\circ$.

\end{enumerate}

In the next step, after finding the beginning of a meteor, we computed the
length of the meteor path. For this purpose we used equations given in
Rendtel et al. (1995). Knowing the radiant position, the beginning of the
meteor path and its length we were able to find the equatorial
coordinates of the end of the meteor.

\subsection{Introducing the errors}

During a real observation no one is capable of exactly determining the
meteor's path and velocity. Thus in an artificial database we have to
modify the paths and velocities of the events introducing the errors
caused by the observers.

There are two error components which affect the direction and position
of the meteor path. They are a tilt $\epsilon$ and a parallel shift $d$
as is shown in Fig. 4. An analysis of these quantities in visual
observations made by experienced observers was done by Koschack (1991).
Their distributions, according to that paper, are shown in Fig. 5. The
solid lines correspond to the Gaussian fits described by the parameters
given also in Fig. 5.


   \begin{figure}[h]
   \centering
\includegraphics[scale=.55]{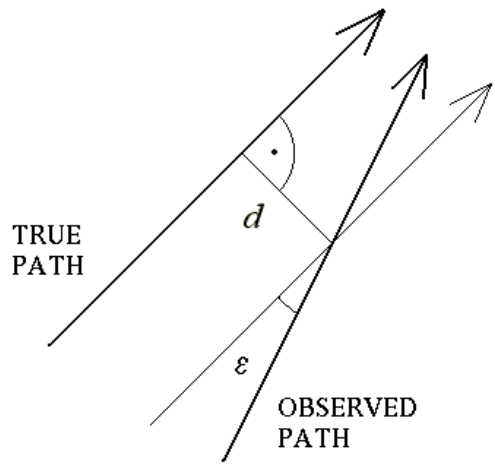}
      \caption{The plotting errors a tilt $\epsilon$ and a parallel shift $d$
              }
         \label{FigVibStab}
   \end{figure}

There are two error components which affect the direction and position
of the meteor path. They are a tilt $\epsilon$ and a parallel shift $d$ 
as is shown in Fig. 4. An analysis of these quantities in visual
observations made by experienced observers was done by Koschack (1991).
Their distributions, according to that paper, are shown in Fig. 5. The
solid lines correspond to the Gaussian fits described by the parameters
given also in Fig. 5.

We used these fits for randomly drawing the tilt and shift for each meteor
and modifying its path.

Due to the errors made by the observer the path of the meteor is not
only tilted and shifted. Additionally observers often change the length
of the meteor path, plotting it as shorter or longer in comparison with
its real length.

In our approach we calculated the observed length of the meteor path $l$
and modified its beginning and end using the Gaussian distributions with
a mean value equal to zero and a standard deviation equal to $l/10$. 

More than just the direction and length of the meteor are modified due
to the errors inputted by the observer. Another factor which is randomly
changed during an observation is the meteor's angular velocity. Knowing
the entry velocity $V_\infty$ of the event and its location on the   
celestial sphere we can calculate its theoretical angular velocity.

As was shown by Koschack (1991) the error in angular velocity inputted
by the observer depends on the angular velocity itself. The distribution
of these errors in different ranges of velocities is shown in Fig. 6.
Each of these distributions was fitted with a Gaussian function shown as
solid line. The mean values of these Gaussian functions were assumed as
zero and the obtained standard deviations are given in each panel.


   \begin{figure}[h]
   \centering
\includegraphics[scale=.6]{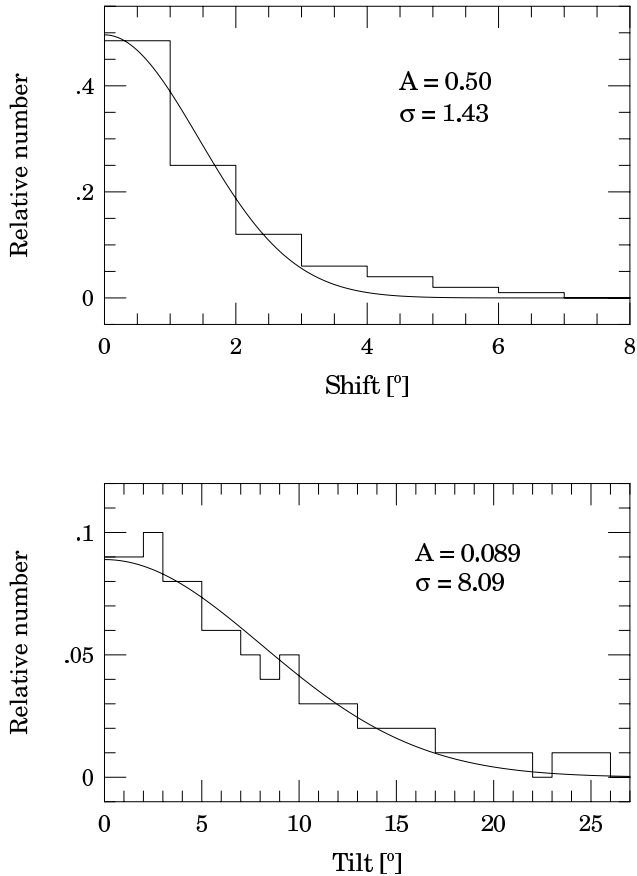}
      \caption{The distribution of tilts and shifts obtained from
observations by experienced observers according to Koschack (1991).
Solid lines denote the Gaussian functions with amplitude and standard
deviation given in each panel.
              }  
         \label{FigVibStab}
   \end{figure}

These distributions were used for randomly modifying the angular
velocities of the meteors in our artificial sample.

As an example of a result, in Fig. 7 and Fig. 8, we show different kinds
of maps produced by {\sc radiant} software for our sample of 2465
Perseids. Fig. 7 shows results obtained for paths of the Perseids before
introducing the errors made by the observer. In the upper panel of this
figure we show the paths of the meteors in the sky. The map is centered
at the radiant of the Perseids for $\lambda_\odot=125^\circ$. The
largest circle has a radius equal to $90^\circ$. One can note that some
meteors do not radiate exactly from the center of the picture. This is
caused by the fact that the radiant of the Perseids moves across the
celestial sphere and meteors observed at the middle of July radiate from
another point than meteors noted one or two weeks later.

In the middle panel of Fig. 7 we show an intersection map obtained using
the tracings mode of the {\sc radiant} software. Due to the assumed
almost one degree radius of the radiant, not all meteors intersect in
the center of the radiant. The elongated shape of the radiant is caused
by the distribution of the meteors which are observed mostly at the
western side of the radiant at this time of the year.

         
   \begin{figure}[h]
   \centering
\includegraphics[scale=.6]{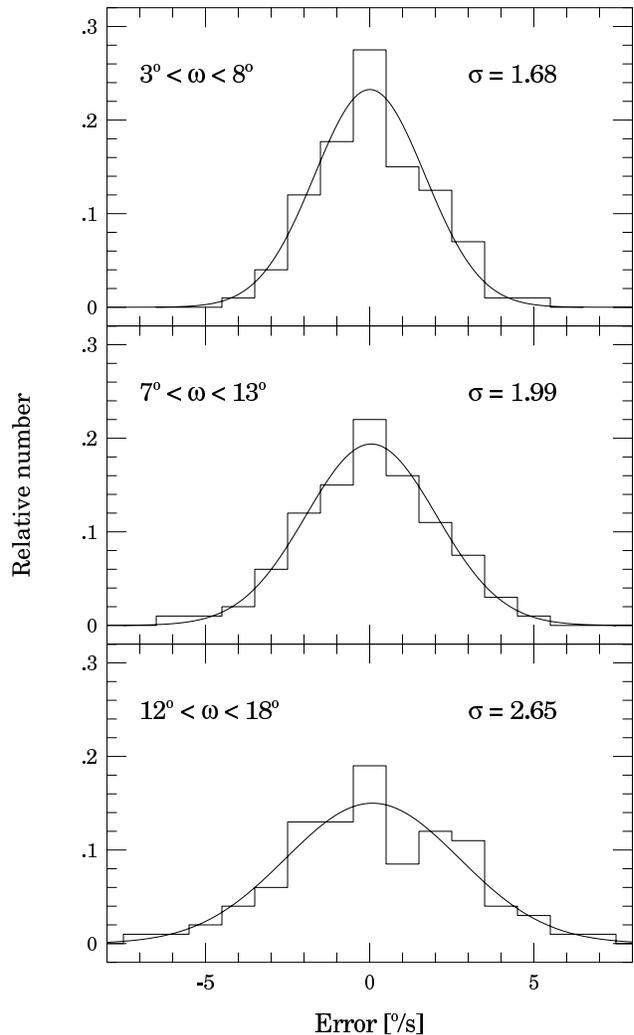}
      \caption{The distribution of errors in angular velocities for
observations of experienced observers according to Koschack (1991).
Solid lines denote the Gaussian functions with a standard   
deviation given in each panel.
              }
         \label{FigVibStab}
   \end{figure}

In the lower panel of Fig. 7 we show the PPR map for our artificial
sample of Perseids. It is clear that the radiant is compact and circular
and its position is correct.

The same analysis was made for the Perseids with meteor paths and
velocities randomly changed using the observational errors and the
result is shown in three panels of Fig. 8. At the upper panel one can
notice that meteors do not always radiate exactly from the center of the
radiant.

\clearpage


   \begin{figure}[h]
   \centering
\includegraphics[scale=.5]{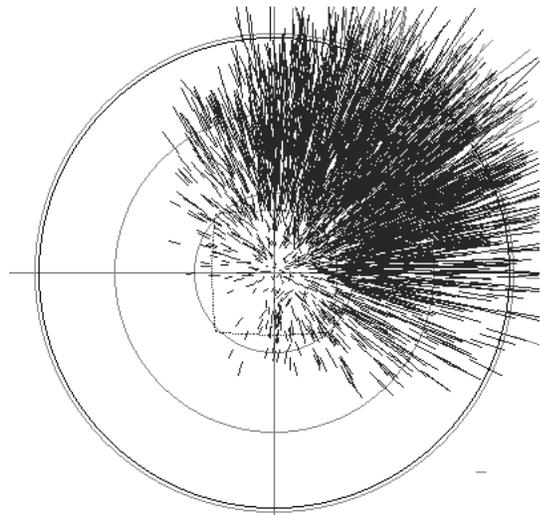}
\includegraphics[scale=.5]{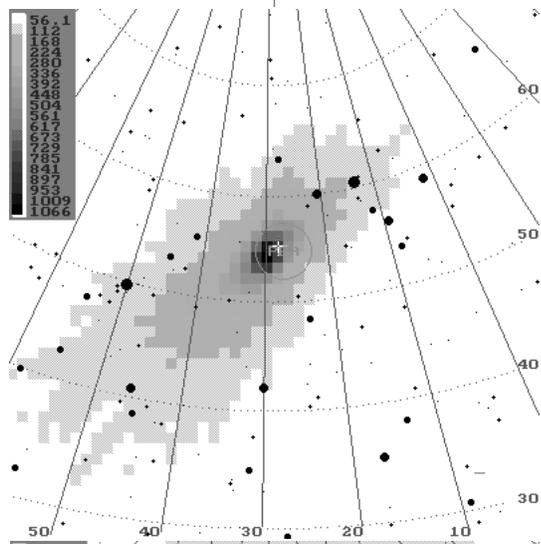}
\includegraphics[scale=.5]{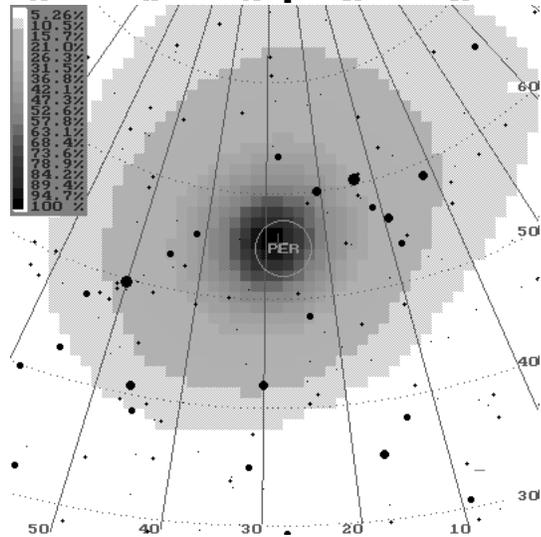}
      \caption{{\sl Upper panel:} Sample distribution of the Perseid shower
meteor paths (observational errors not included
yet). The center of the figure is at the radiant of the Perseids. {\sl
Middle panel:} The radiant of the Perseids obtained using the tracings
method of
the {\sc radiant} software. {\sl Lower panel:} PPR map of the Perseids.
              }
         \label{FigVibStab}
   \end{figure}


   \begin{figure}[h]
   \centering
\includegraphics[scale=.5]{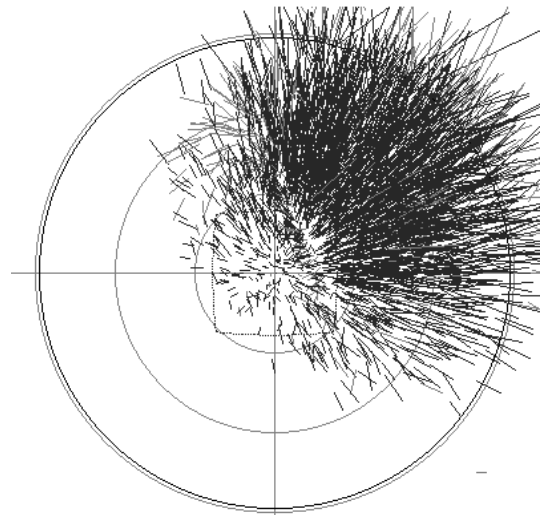}
\includegraphics[scale=.5]{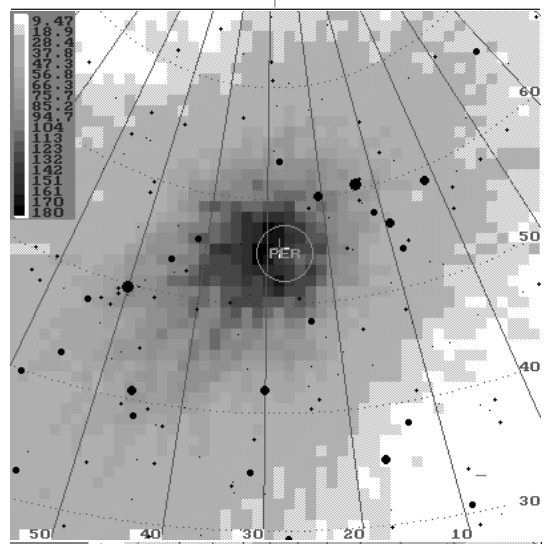}
\includegraphics[scale=.5]{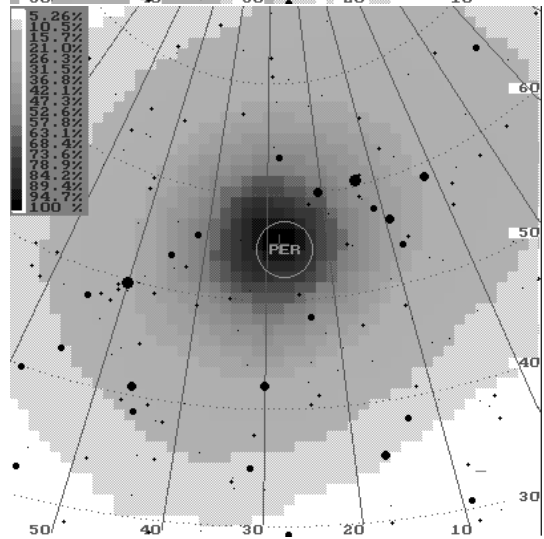}
      \caption{{\sl Upper panel:} Sample distribution of the Perseid shower
meteor paths with observational
errors included. The center of the figure is at the radiant of the Perseids.
{\sl Middle panel:} The radiant of the Perseids obtained using the tracings
method of the {\sc radiant} software. {\sl Lower panel:} PPR map of Perseids.
              }
         \label{FigVibStab}
   \end{figure}

\clearpage

Although the radiant obtained using the tracings method, shown in the
middle panel, is still very clear it is also more diffuse in comparison
with the middle panel of Fig. 7. It is worth noting that there is a
different scale used in both middle panels. In Fig. 7 the highest number
of intersections detected in the center of the radiant is 1066 and in
Fig. 8 it reaches only 180.

The PPR map obtained for the sample of Perseids with introduced
observational errors shows a radiant which is still circular and at the
correct position but significantly more diffuse than in previous case.

\section{Distribution of sporadic meteors over the celestial sphere}

Since the 1950s we know that sporadic meteor radiants are not
distributed uniformly over the celestial sphere, but they are
concentrated in particular regions which are similar to radiants with
radii slightly larger than $20^\circ$ and positions approximately fixed
relative to the Sun. The first three sources associated with the
ecliptic plane were discovered by Hawkins (1957). According to the
latest papers (Jones \& Brown 1993, Brown \& Jones 1995, Poole 1997) we
now identify six such sources. They are antihelion, helion, northern and
southern toroidal centers and also the northern and southern apex.

When constructing the sample database of meteors we have taken into
account the above mentioned sources. We decided to omit the helion, southern
toroidal center and the southern apex. These radiants in July at Polish
geographical coordinates are always either very close to the Sun or below
the horizon. The remaining three sources were included into the database
and their properties taken from Jones \& Brown (1993) are summarized in
Table 4.

\begin{table}[h] 
\caption{The ecliptic coordinates and radii of the
sporadic meteor sources included in our artificial database}
\begin{center}
\begin{tabular}{|l|c|c|c|} 
\hline 
\hline 
Source & $\lambda-\lambda_\odot$ & $\beta$ & $R$ [$^\circ$]\\ 
\hline 
\hline
Antihelion & 198$^\circ$ & $0^\circ$ & 18$^\circ$ \\
Northern Toroidal Source & 271$^\circ$ & $+57^\circ$ & 19$^\circ$  \\
Norther Apex & 271$^\circ$ & $+19^\circ$ & 21$^\circ$ \\
\hline
\hline
\end{tabular}
\end{center}
\end{table}

According to the work of Jones \& Brown (1993) and Poole (1997) all sources
seem to have similar activity. Thus we assumed that the number of the
sporadic events radiating from each center is proportional only to the
sine function of the center altitude. The whole number of sporadics observed
during each hour was divided between these sources. Especially in the
evening hours, when all centers are either below the horizon or only
slightly over it, we assumed the existence of another source. It was
centered at the zenith and its radiant had a radius equal to
60$^\circ$.

Knowing the positions and the radii of all four sources, for the purpose
of finding the location of sporadic meteor paths, we followed the
procedures applied for shower meteors described in section 4.1.

Also the procedures for introducing the errors into the meteor paths was
the same as described in section 4.2.

The sporadic sources, contrary to the shower radiants, are not
characterized by meteors with a common entry velocity. Thus for each
sporadic source we used the entry velocity distributions given by Jones
\& Brown (1993) (see their Figs. 7, 8, and 9). These velocities were
also changed according to the error distributions presented in Fig. 6.

\section{$\alpha$-Cygnids}

The $\alpha$-Cygnid shower was discovered by W.F. Denning (1919). After
his observations we have rather poor information about the activity of
this shower. A reasonable determination of the position of the radiant
was done using photographic observations, and based on the one captured
event, the equatorial coordinates of the radiant were
$\alpha=304.5^\circ$ and $\delta=+48.7^\circ$ with a geocentric velocity
equal to $V_\infty=41.0$ km/s (Babadzhanov \& Kramer, 1961)

The first comprehensive study of the $\alpha$-Cygnids, based on the 11
years of visual observations, was presented by Jenniskens (1994).
According to this work the $\alpha$-Cygnids are a weak shower with maximum
ZHRs equal to $2.5\pm0.8$ occurring at a solar longitude
$\lambda_\odot=116.0^\circ$. The meteors from this shower are
observed from $\lambda_\odot=105^\circ$ to $\lambda_\odot=127^\circ$.

In recent years the $\alpha$-Cygnid shower was intensively analyzed by the
Polish {\sl Comets and Meteors Workshop} (Olech et al. 1999a, 1999b,
Stelmach \& Olech, 2000). Here we only briefly mention the results
presented in the latest of these papers, which is based on the most
comprehensive sample. According to Stelmach \& Olech (2000) the radiant
of the $\alpha$-Cygnids is at $\alpha=305^\circ$ and $\delta=+45^\circ$.
The activity of this shower lasts from the beginning to the end of July with
a quite obvious maximum at $\lambda_\odot=114.8^\circ\pm0.5^\circ$. Maximal
ZHRs are equal to $2.4\pm0.1$.

As was pointed out by Olech et al. (1999a) the $\alpha$-Cygnid
shower was probably not recognized before the study of Jenniskens (1994)
because of a lack of data in photographic meteor databases around the
peak date of the shower. 

Stelmach \& Olech (2000) presented only the preliminary results for year
1999. As the {\it CMW} database from period 1996-1999 is complete now,
we decided to recalculate PPR maps for $\alpha$-Cygnids. We selected
6772 meteors observed in period June 30 - July 31. The PPR maps were
centered at $\alpha=303^\circ$ and $\delta=+45^\circ$. According to
earlier results, computations were performed for the moment of
$\lambda_{\odot (max)}=115^\circ$ with daily drift of the radiant equal
to $\Delta\lambda=+1.0^\circ$. The assumed entry velocity was 
$V_\infty=41$ km/s. From our sample we excluded meteors slower than
$1^\circ/sec$ and faster than $30^\circ/sec$.

\clearpage


   \begin{figure}[h]
   \centering
\includegraphics[scale=.6]{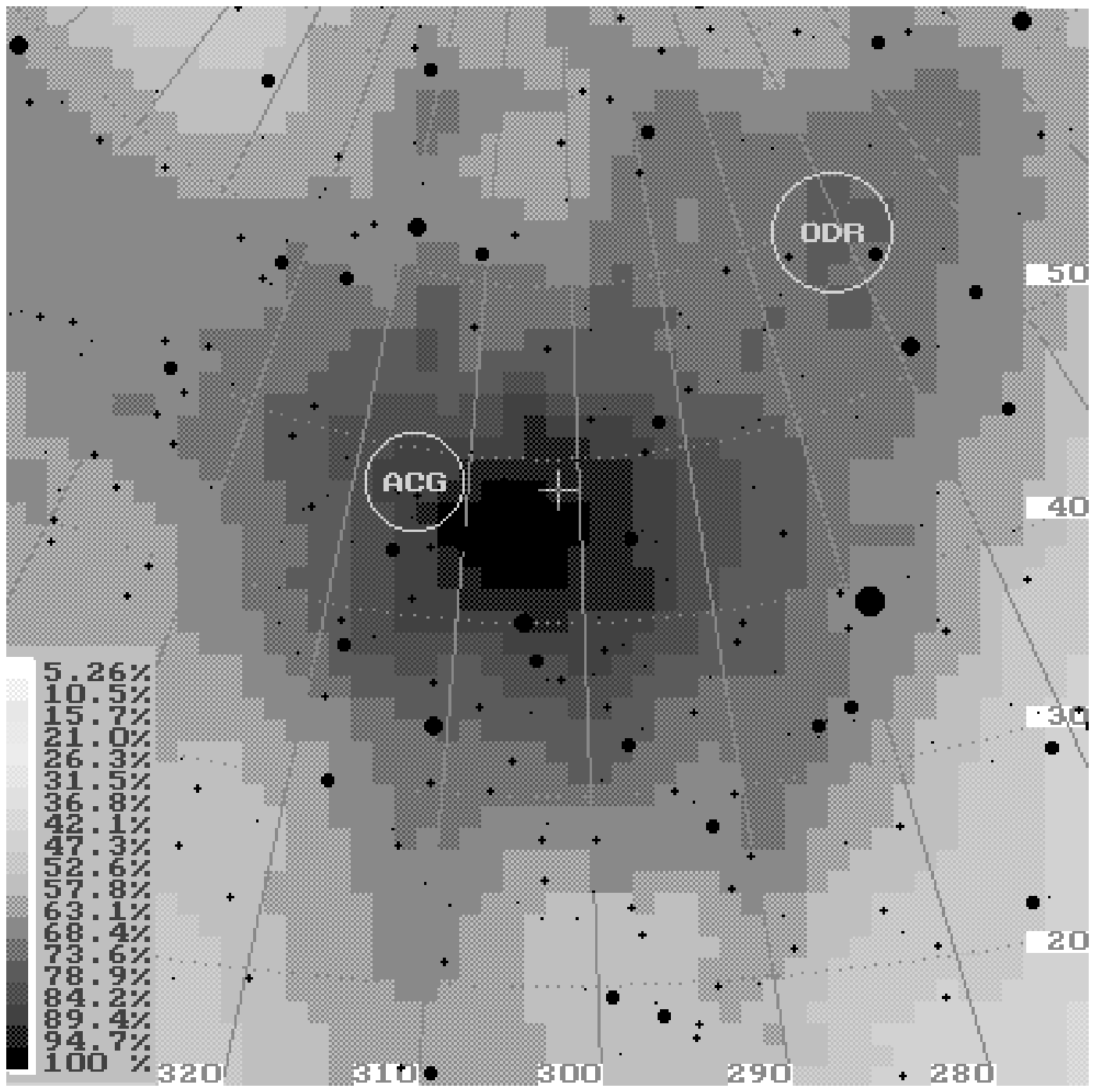}
\includegraphics[scale=.6]{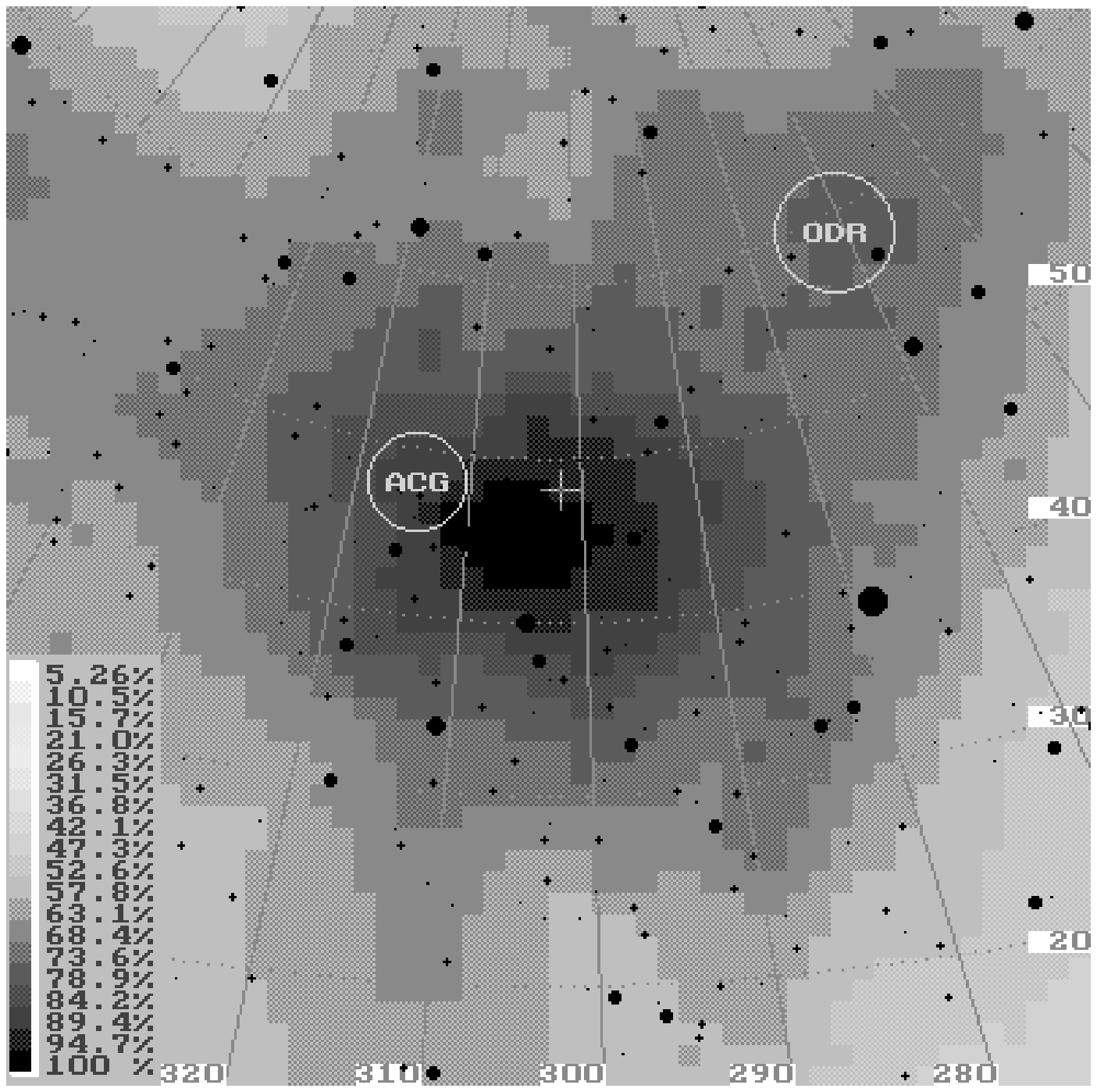}
      \caption{PPR maps for a real sample of 6772 meteors observed in
years 1996-1999. All maps are computed for the following parameters:
$\lambda_\odot=115^\circ$, $\Delta\lambda=1.0^\circ$ and $V_\infty=41$
km/s. The maximum distance of the meteor from the radiant is 50 and
$85^\circ$, respectively from the upper to lower panel. 
              }
         \label{FigVibStab}
   \end{figure}




   \begin{figure}[h]
   \centering
\includegraphics[scale=.6]{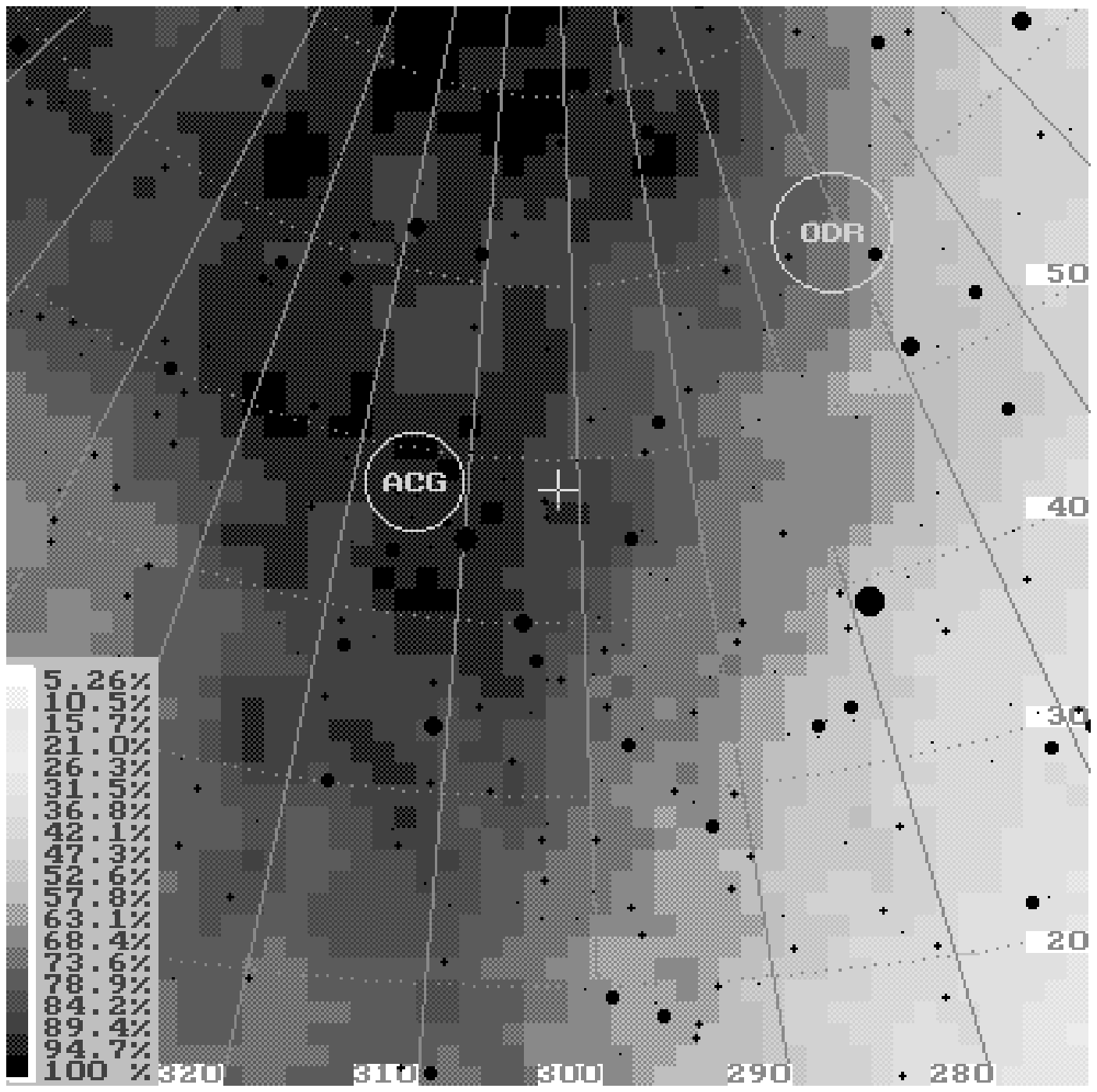}
\includegraphics[scale=.6]{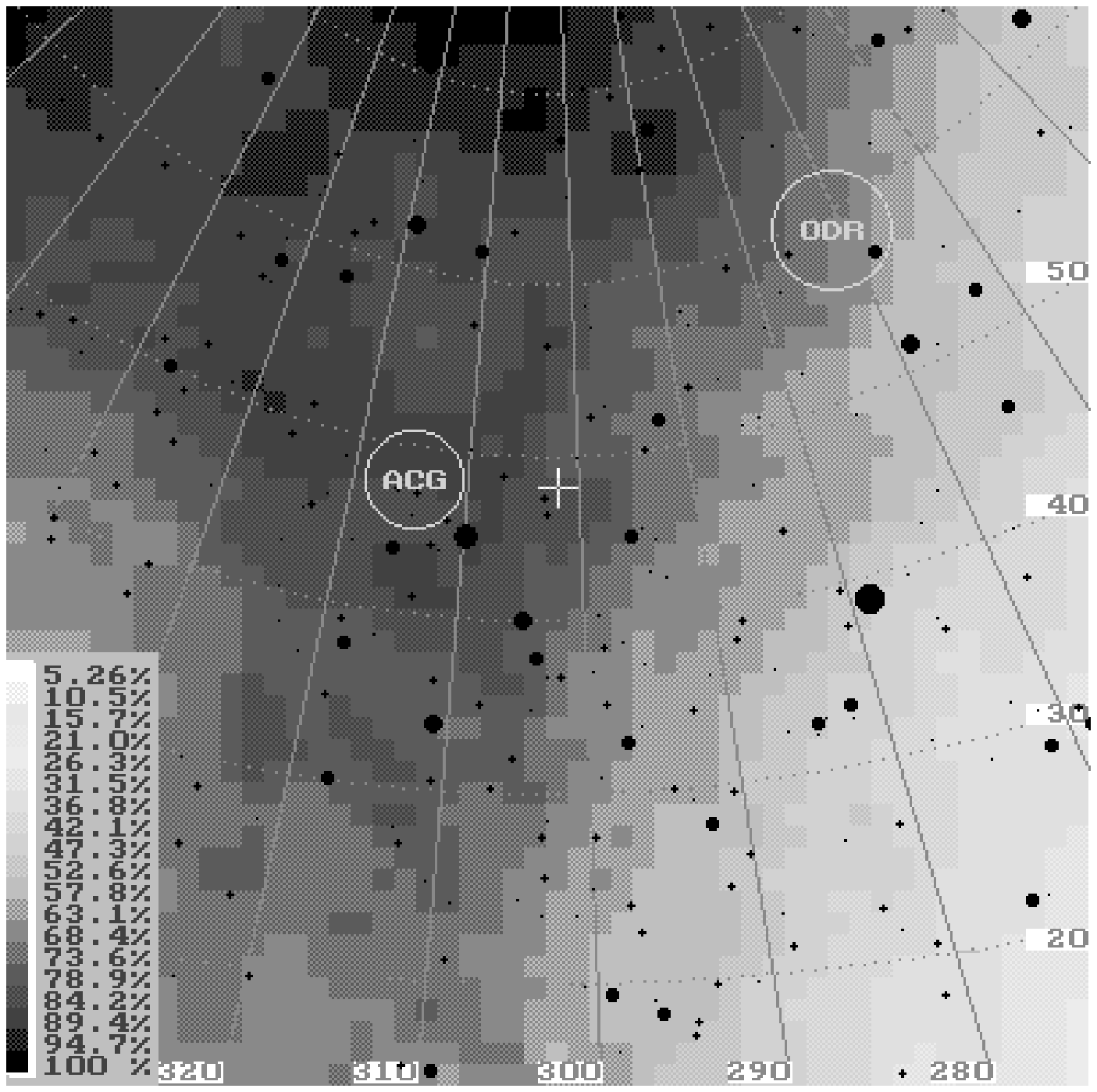}
      \caption{PPR maps for an artificial sample of 21516 meteors.
All maps are computed for the following parameters:
$\lambda_\odot=115^\circ$, $\Delta\lambda=1.0^\circ$ and $V_\infty=41$
km/s. The maximum distance of the meteor from the radiant is 50 and
$85^\circ$, respectively from the upper to lower panel.
              }
         \label{FigVibStab}
   \end{figure}

\clearpage


   \begin{figure}[h]
   \centering
\includegraphics[scale=.6]{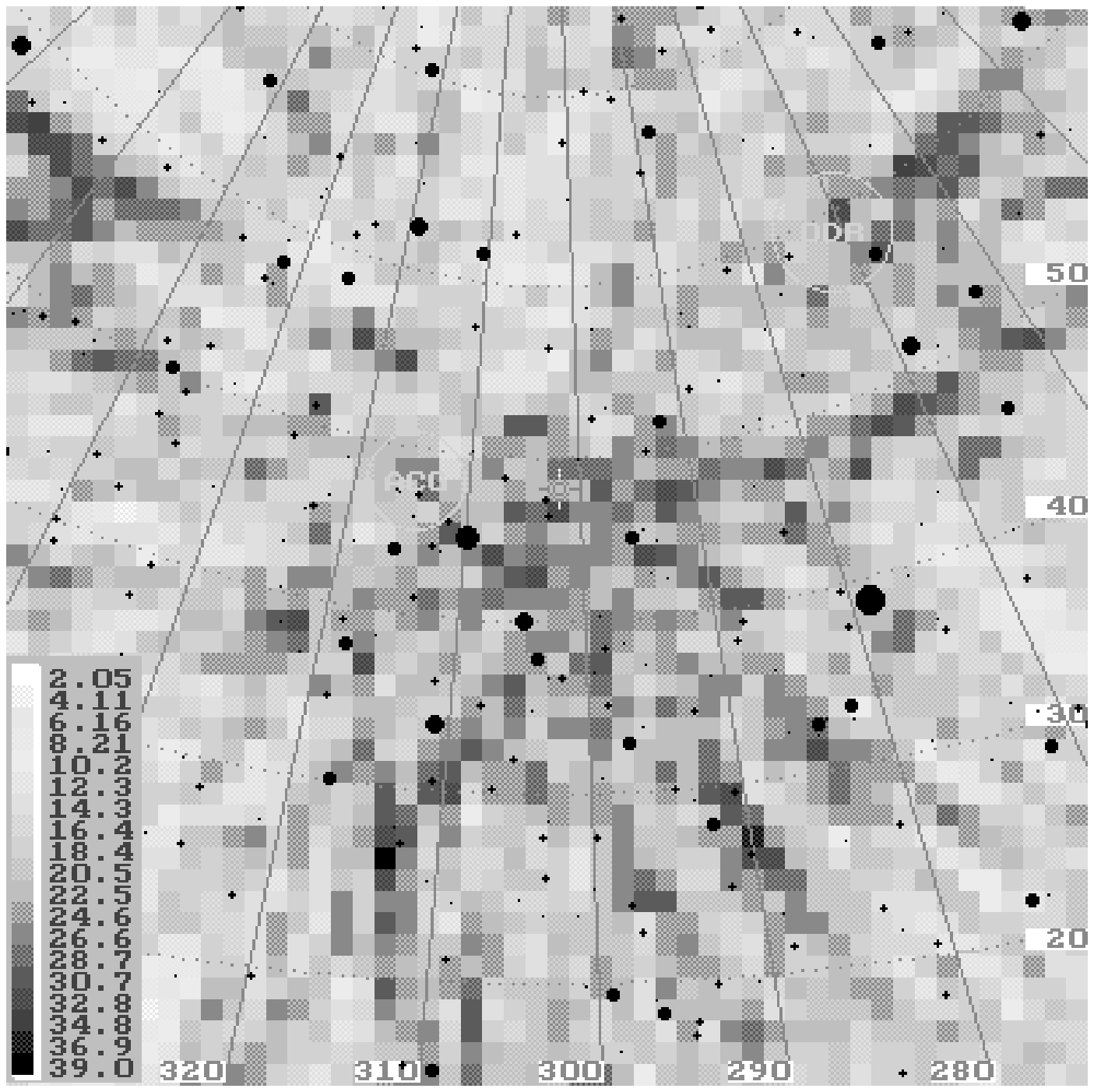}
\includegraphics[scale=.6]{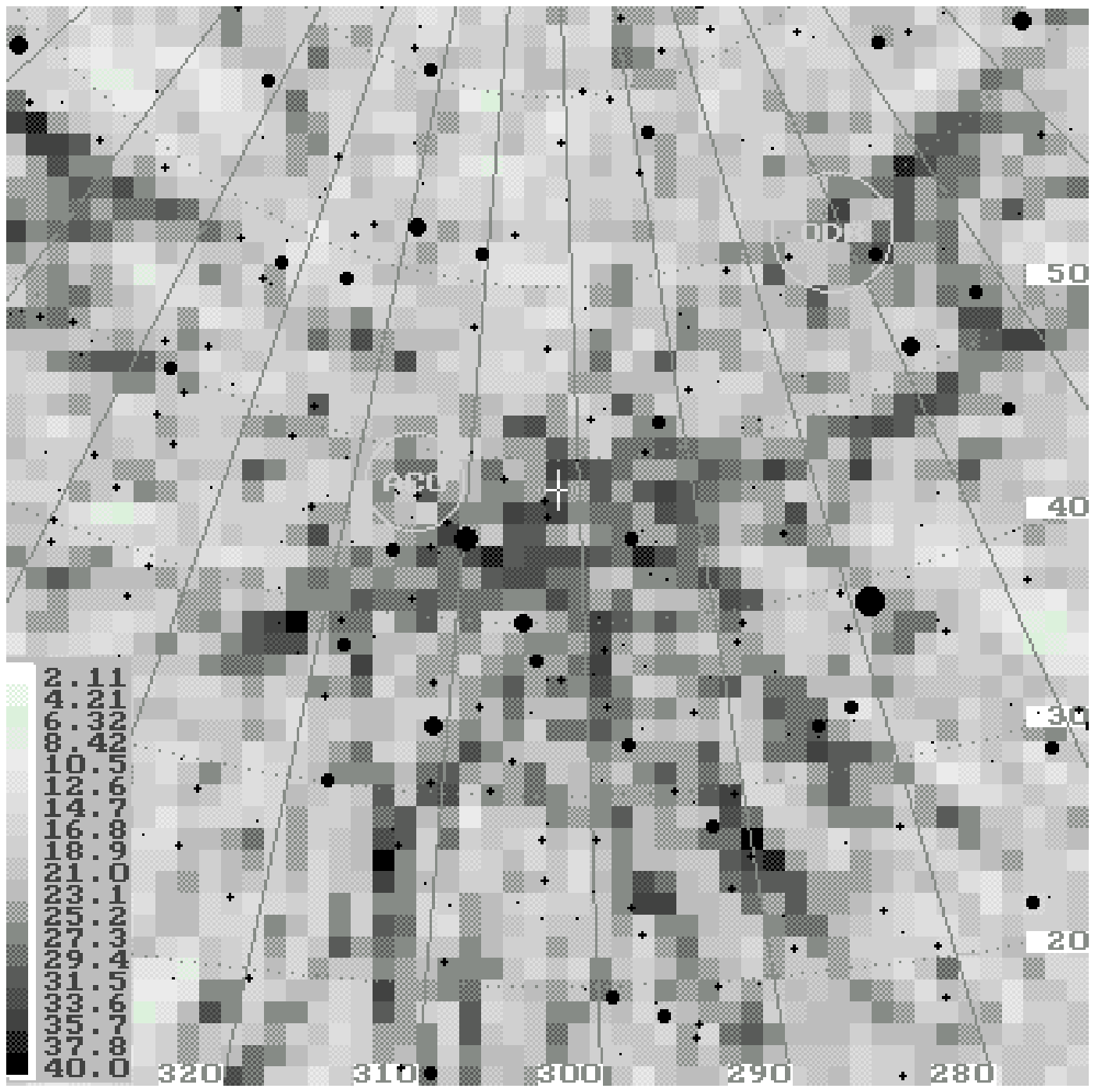}
      \caption{Tracings maps for a real sample of 6772 meteors observed in
the years 1996-1999. All maps are computed for the following parameters:
$\lambda_\odot=115^\circ$, $\Delta\lambda=1.0^\circ$ and $V_\infty=41$
km/s. The maximum distance of the meteor from the radiant is 50 and
$85^\circ$, respectively from the upper to lower panel.
              }
         \label{FigVibStab}
   \end{figure}

               

   \begin{figure}[h]
   \centering
\includegraphics[scale=.6]{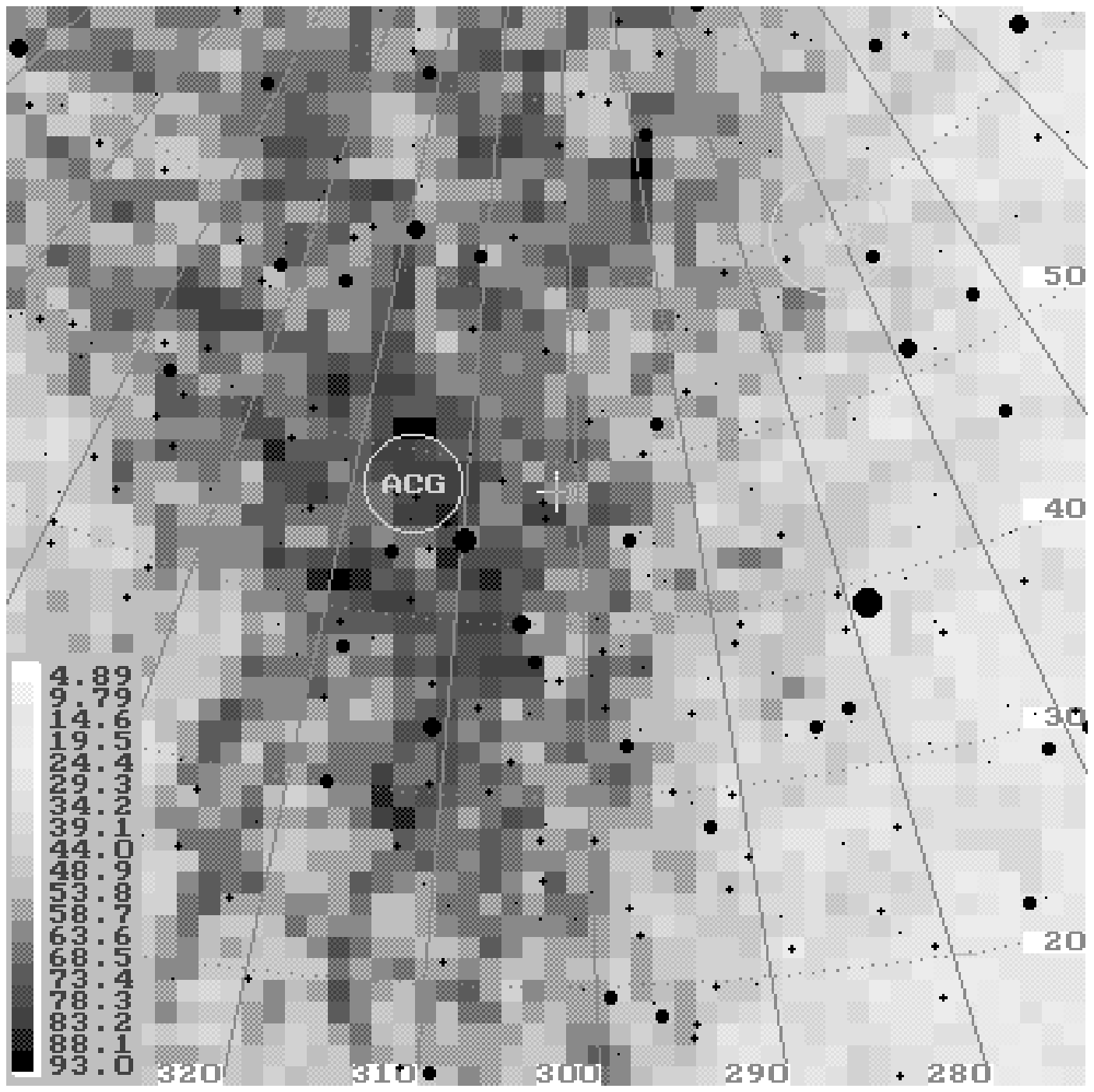}
\includegraphics[scale=.6]{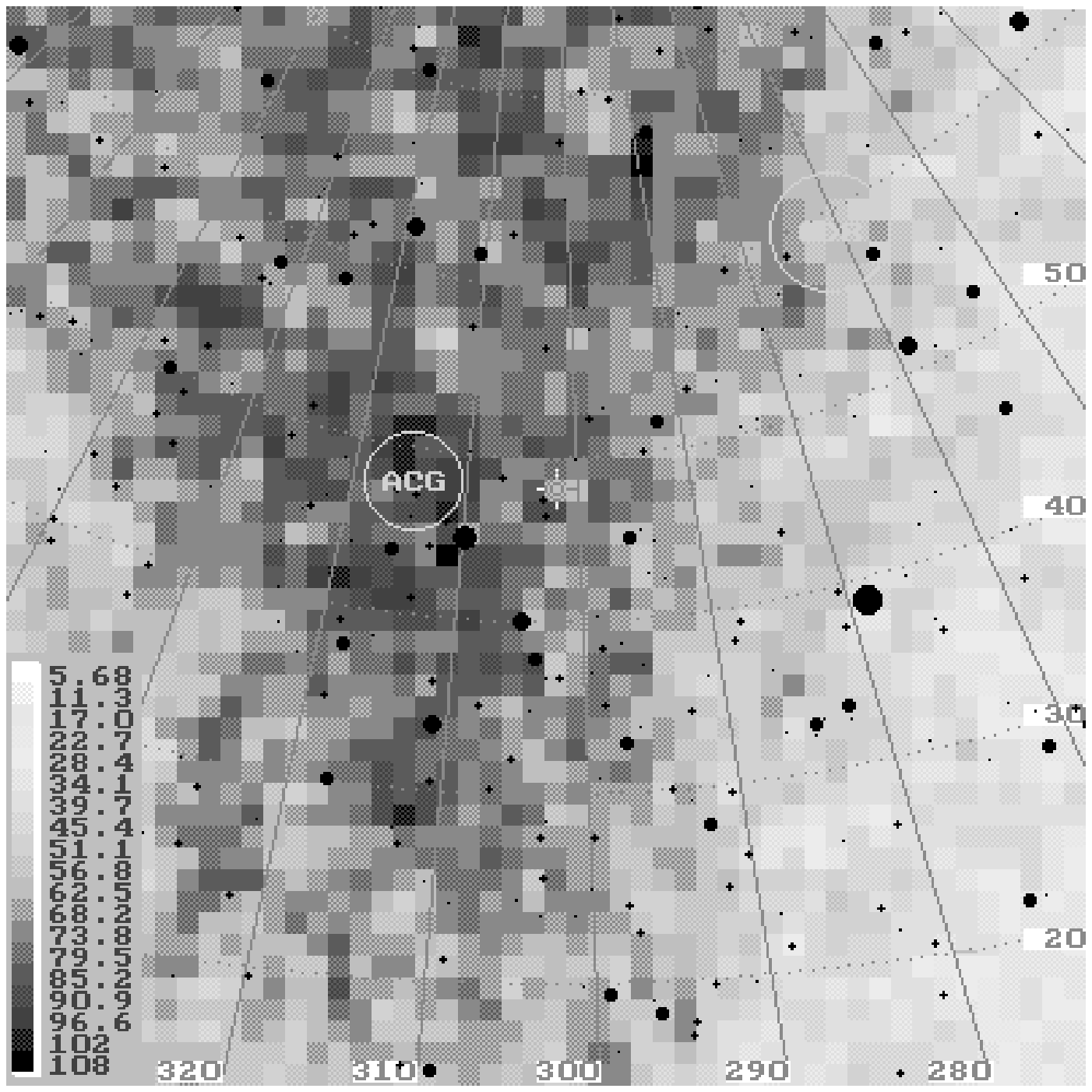}
      \caption{Tracings maps for an artificial sample of 21516 meteors.
All maps are computed for the following parameters:
$\lambda_\odot=115^\circ$, $\Delta\lambda=1.0^\circ$ and $V_\infty=41$
km/s. The maximum distance of the meteor from the radiant is 50 and 
$85^\circ$, respectively from the upper to lower panel.
              }
         \label{FigVibStab}
   \end{figure}

\clearpage

\noindent The calculations were done rejecting meteors placed at
distances larger than 50 and 85 degrees from the radiant. The final
results are shown in Fig. 9.

One can clearly detect the circular radiant close to the center of each
map. The {\sc radiant} software allows two ways of estimating the
radiant position from the PPR map. The first of them uses the framed
part of a PPR map and computes the simple mean position of the radiant
weighted by the values of computed probability. The second one also uses
the framed part of a PPR map but in this case the probability
distribution is fitted with a two-dimensional Gaussian function. These
options were used for each of our three PPR maps resulting with six
radiant position determinations. The simple mean of these values is
$\alpha=303.9^\circ\pm0.5^\circ$ and $\delta=+45.3^\circ\pm0.6^\circ$.
The quoted errors are simple standard deviations of the mean and due to
the fact that our six determinations are not completely independent, the
real errors might be from two to three times larger.

A similar approach was undertaken using the artificial sample and the
results are shown in three panels of Fig. 10, where we presented the PPR
maps calculated rejecting meteors placed at distances larger than 50
and 85 degrees from the suspected position of the radiant. There is no
trace of any circular structure as was detected in the case of real meteors.
High probabilities (black areas) at these figures are caused by two
reasons. First, the PPR map is always scaled to the highest probability
point and its value is assumed to be 100\%. Taking into account that we
assumed the diffuse sporadic source placed at a zenith with a radius equal
to $60^\circ$, we should expect the presence of this feature in our maps.
The zenith, at Polish latitudes in July, lies near Deneb ($\alpha$ Cyg)
thus we are not surprised by the high probabilities of detecting the meteors
radiating from this region of the sky. But, as we mentioned, the PPR
map for the artificial sample does not show any circular radiant as in
the case of the real sample, strongly suggesting that the $\alpha$-Cygnids
are the real shower.

Our conclusions are confirmed by a comparison of the real and artificial
sample maps obtained by the tracings method of the {\sc radiant}
software. At Fig. 11 we have presented tracings maps for real sample
meteors placed respectively within 50 and 85 degrees from the center
of the map. In all three cases one can see a clear enhancement of
the intersections (about 30-40) at the radiant of the $\alpha$-Cygnids. No
such picture is present in the case of the artificial sample for which the
results are shown in Fig. 12. Now, the distribution of intersections is more
diffuse and is centered mostly at the zenith rather than at the center
of the map.

Finally, we conclude that we are unable to reproduce the circular and
high quality picture of the radiant of the $\alpha$-Cygnids using the
artificial sample. Such a structure is clearly detected in the real database
strongly suggesting that the $\alpha$-Cygnid shower indeed exists.

\section{Delphinids}

The Delphinids are not a new shower. The existence of meteors radiating
from the constellation of Delphinus was suggested by Russian and Polish
meteor sources (Abalakin 1981, Kosinski 1990). In the beginning of the 1990s
the shower was studied by Bulgarian observers (Velkov
1996). The comprehensive analysis of this shower was undertaken by the
{\sl Comets and Meteors Workshop} (Olech et al. 1999b, Wi\'sniewski \&
Olech 2000, 2001). It showed that the Delphinids are a very weak shower
with the maximum at $\lambda_\odot=125^\circ$ with ${\rm
ZHR}=2.2\pm0.2$.

The estimated radiant position was $\alpha=312^\circ$ and
$\delta=+12^\circ$. It is very close to the antihelion source of
sporadic meteors, whose position for $\lambda_\odot=125^\circ$ is
$\alpha=325^\circ$ and $\delta=+14^\circ$. Thus one can presume that the
radiant of the Delphinids obtained by Wi\'sniewski \& Olech (2000, 2001)
is not real but rather comes from crossing the paths of meteors from
the antihelion source and other ecliptic showers active during the second
part of July.

To clarify this situation we decided to reanalyze the real sample of
Wi\'sniewski \& Olech (2001) collected in the years 1996-1999. We selected
6468 meteors observed between July 9th and July 31st and then computed PPR
maps for meteors within 50, 70 and 85 degrees from the radiant. We also
assumed that the entry velocity $V_\infty$ is equal to 35 km/s and
meteors in the sky are slower than $30^\circ/sec$. The assumed daily
drift of the radiant was $\Delta\lambda=1.0^\circ$.

The results of our computation are presented in three panels of Fig. 13.
The upper panel shows the radiant of the Delphinids for meteors closer than
$50^\circ$ from the center of the map. The big cross marks the
position of the antihelion source. As in the case of the $\alpha$-Cygnids we
computed the mean position of the Delphinids' radiant combining the
estimates obtained for different PPR maps shown in Fig. 13. This mean
position, equal to $\alpha=313.4^\circ\pm0.6^\circ$ and
$\delta=+8.6^\circ\pm2.7^\circ$, is marked by the asterisk.

A similar computation was performed for 20376 meteors from the
artificial sample. All parameters were the same as in the case of the real
sample. The results are shown if Fig. 14. Again the big cross denotes
the position of the antihelion source and the asterisk the radiant of the
Delphinids obtained from visual observations of the {\sl CMW}.

A comparison of Fig. 13 and 14 shows clear differences. First, in the
case of the real sample, the radiant of the Delphinids is quite compact
and suffers from intense pollution from ecliptic showers and the
$\alpha$-Cygnids only in PPR maps shown in the middle and lower panels.
The first panel, showing only the closest meteors, gives a quite
circular radiant in a clearly different position from the antihelion
source. All three panels obtained for the artificial sample, at the
position of the radiant of the Delphinids, give the probability of
detecting the radiant around 70\%. This is significantly lower comparing
with almost 100\% probability at the position of the $\delta$-Aquarids N
radiant and also at the antihelion source. High probabilities are




   \begin{figure}[h]
   \centering
\includegraphics[scale=.6]{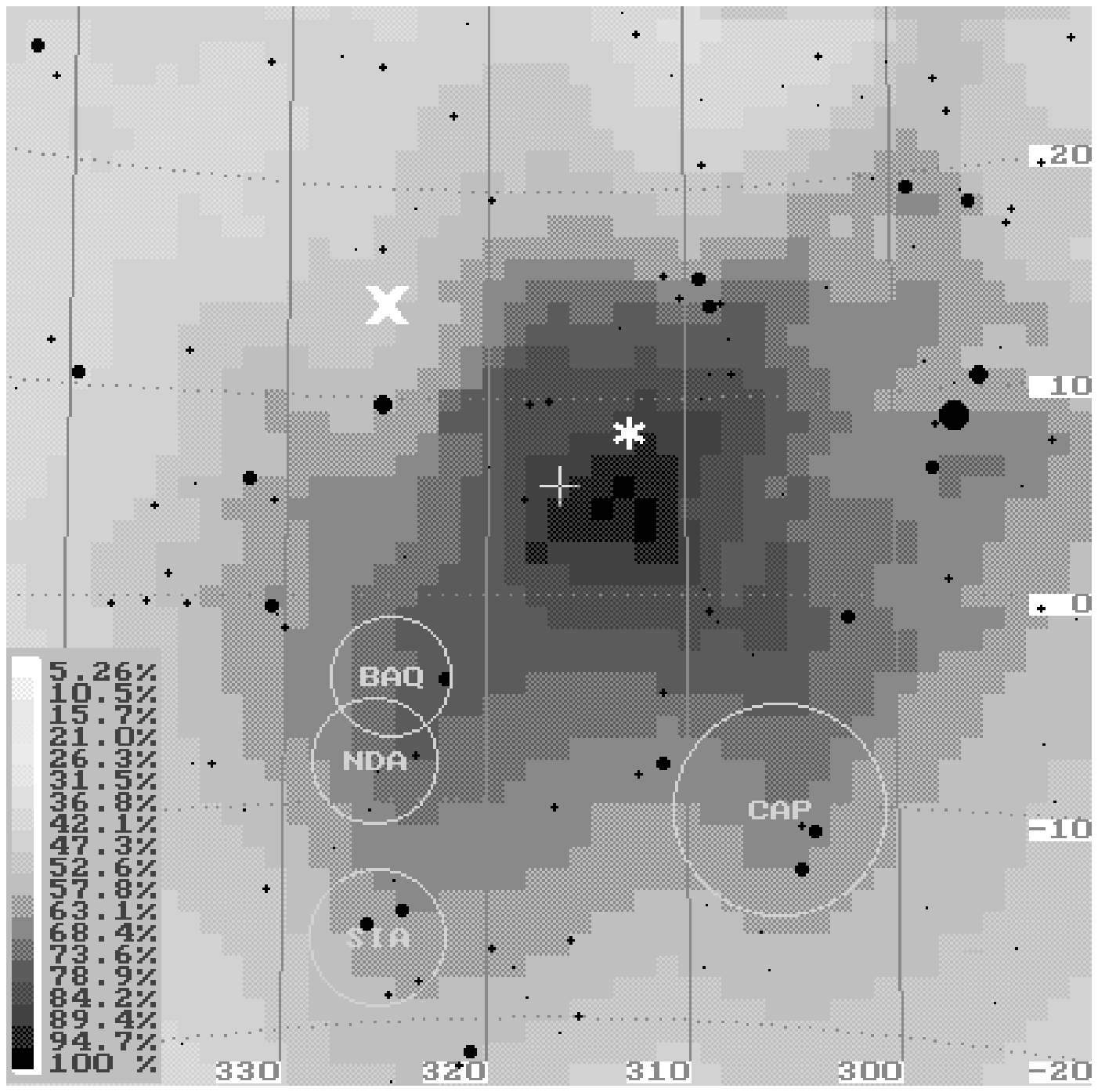}
\includegraphics[scale=.6]{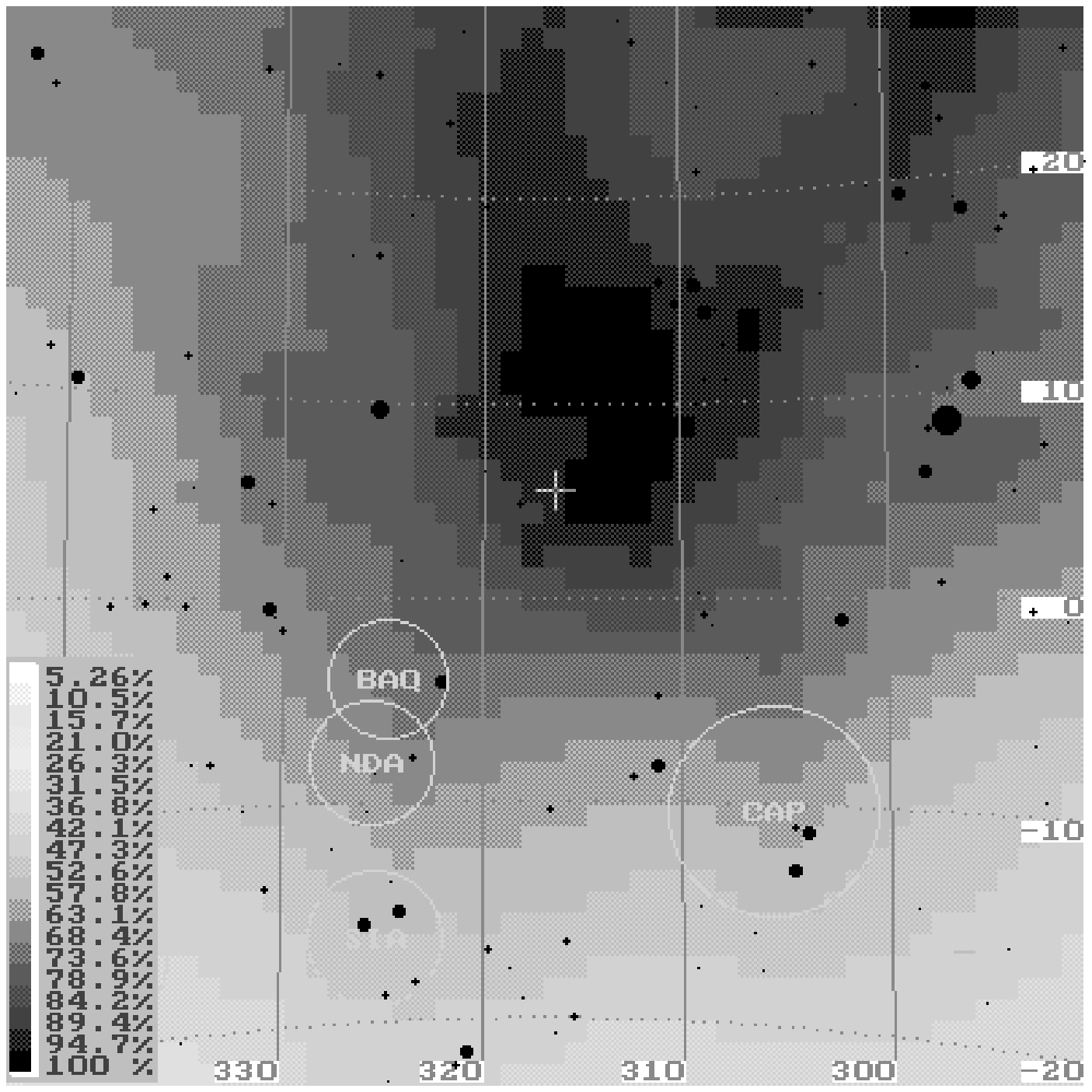}
      \caption{PPR maps for a real sample of 6468 meteors observed in
the years 1996-1999. All maps are computed for the following parameters:
$\lambda_\odot=125^\circ$, $\Delta\lambda=1.0^\circ$ and $V_\infty=35$
km/s. The maximum distance of the meteor from the radiant is 50 and
$85^\circ$, respectively from the upper to lower panel. The big cross
marks the position of antihelion source and the asterisk the mean
position of Delpninids' radiant derived from real observations of {\sl CMW}
              }
         \label{FigVibStab}
   \end{figure}




   \begin{figure}[h]
   \centering
\includegraphics[scale=.6]{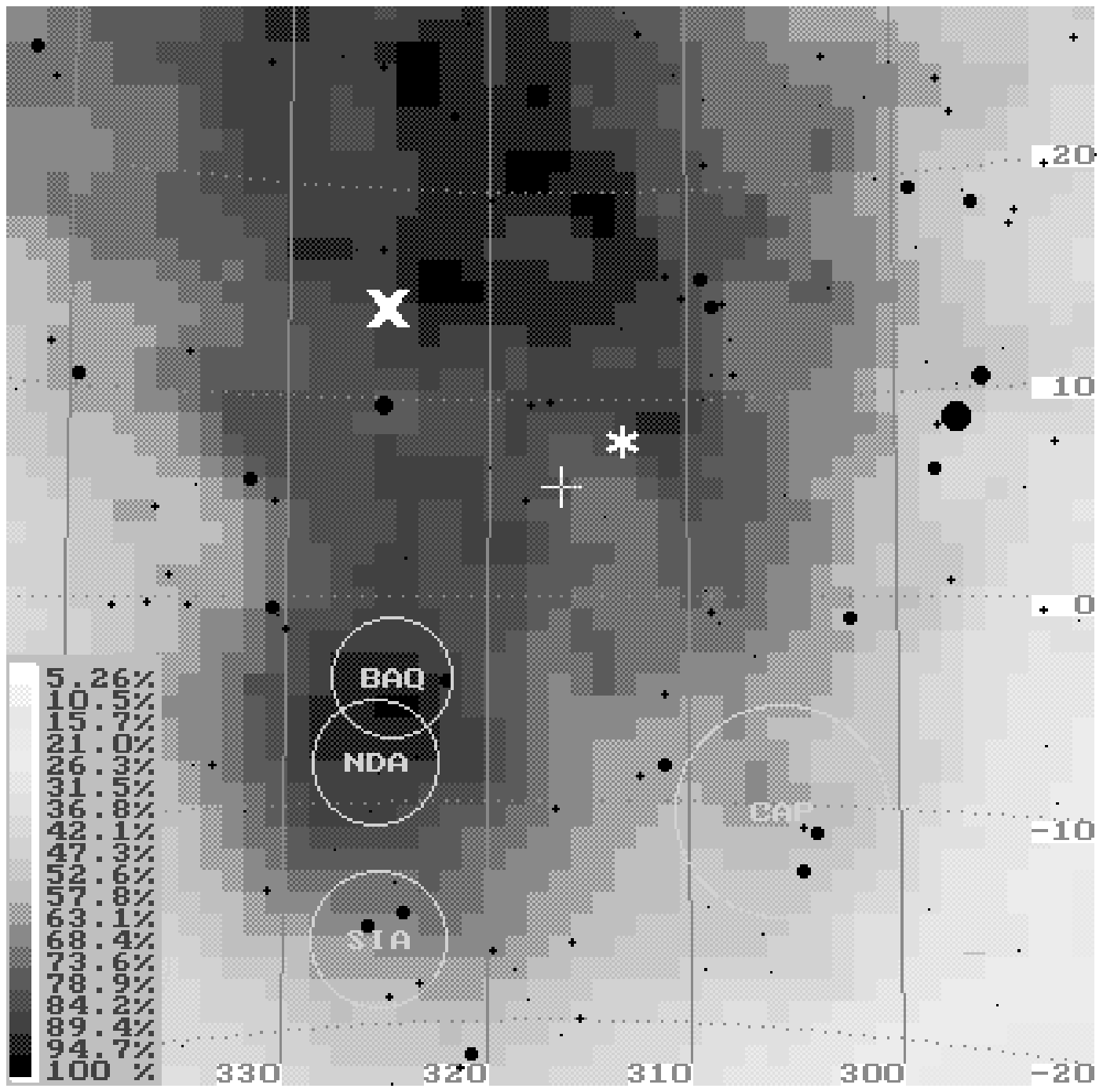}
\includegraphics[scale=.6]{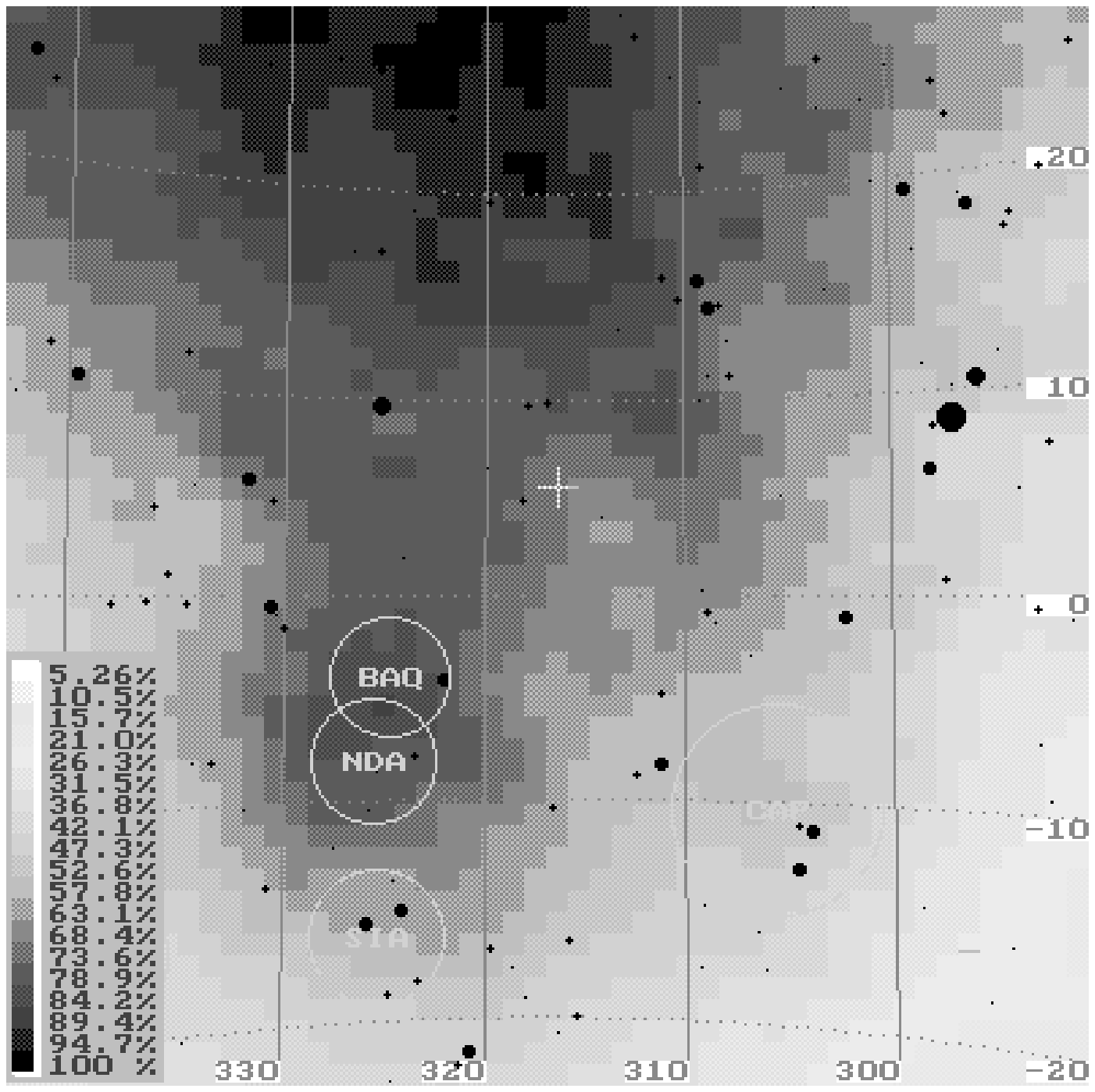}
      \caption{PPR maps for an artificial sample of 20376 meteors observed
in the period July 8-31. All maps are computed for the following parameters:
$\lambda_\odot=125^\circ$, $\Delta\lambda=1.0^\circ$ and $V_\infty=35$
km/s. The maximum distance of the meteor from the radiant is 50 and
$85^\circ$, respectively from the upper to lower panel. The big cross
marks the position of antihelion source and the asterisk the mean
position of Delpninids' radiant derived from real observations of {\sl CMW}
              }
         \label{FigVibStab}
   \end{figure}


   \begin{figure}[h]
   \centering
\includegraphics[scale=.6]{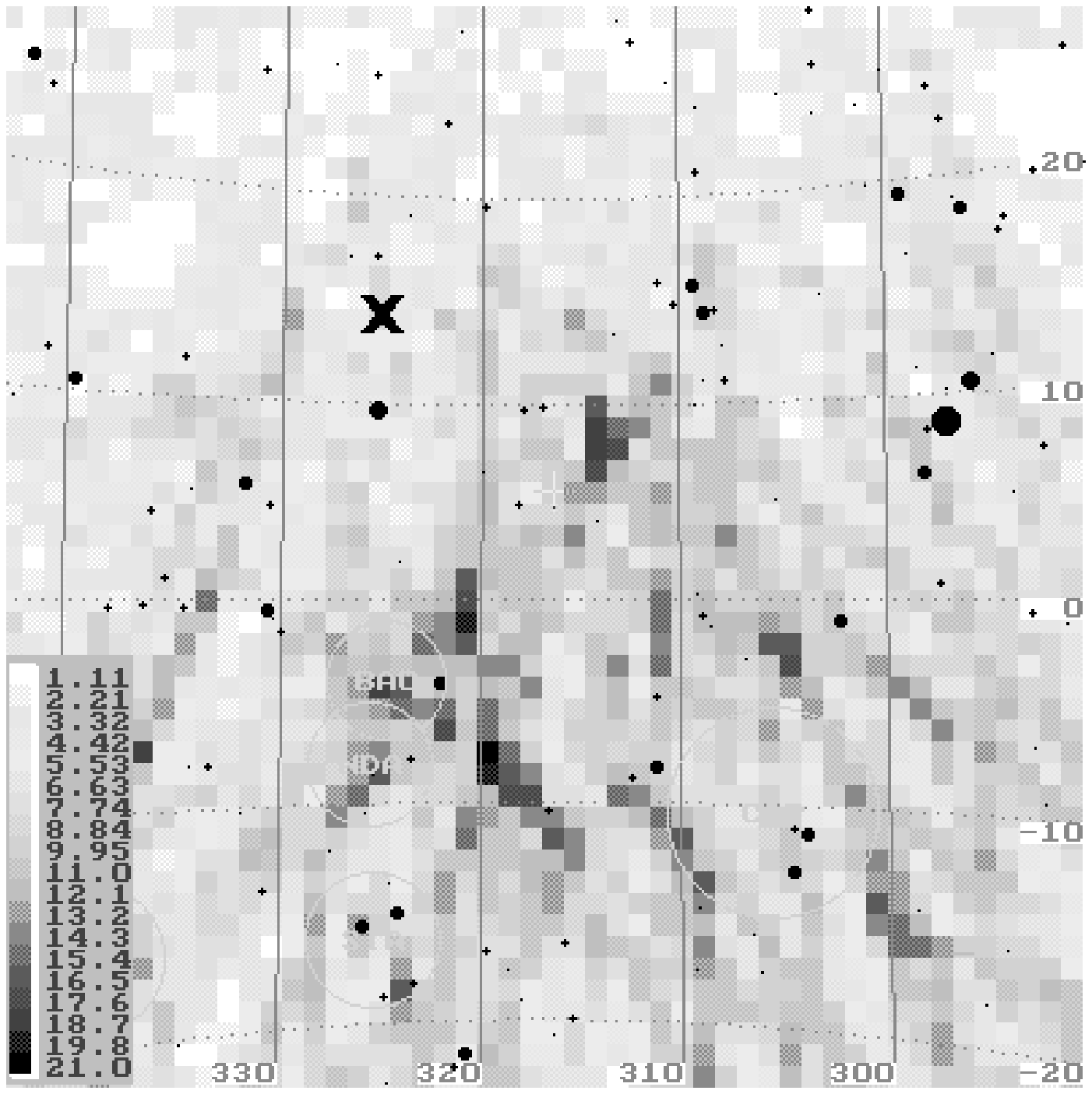}
\includegraphics[scale=.6]{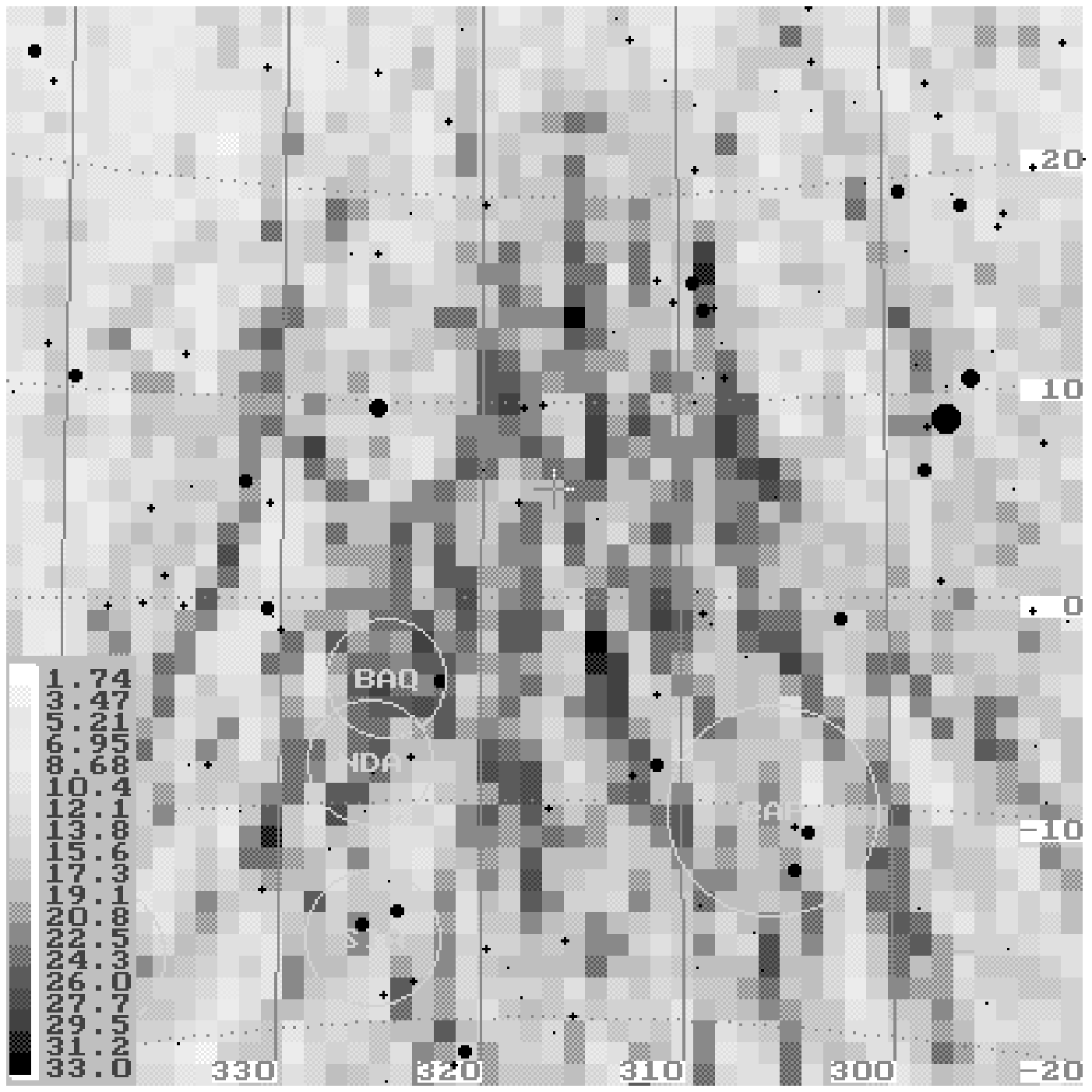}
      \caption{Tracing maps for a real sample of 6468 meteors observed
in the years 1996-1999. All maps are computed for the following parameters:
$\lambda_\odot=125^\circ$, $\Delta\lambda=1.0^\circ$ and $V_\infty=35$
km/s. The maximum distance of the meteor from the radiant is 50 and
$85^\circ$, respectively from the upper to lower panel. The big cross
marks the position of antihelion source.
              }
         \label{FigVibStab}
   \end{figure}

               


   \begin{figure}[h]
   \centering
\includegraphics[scale=.6]{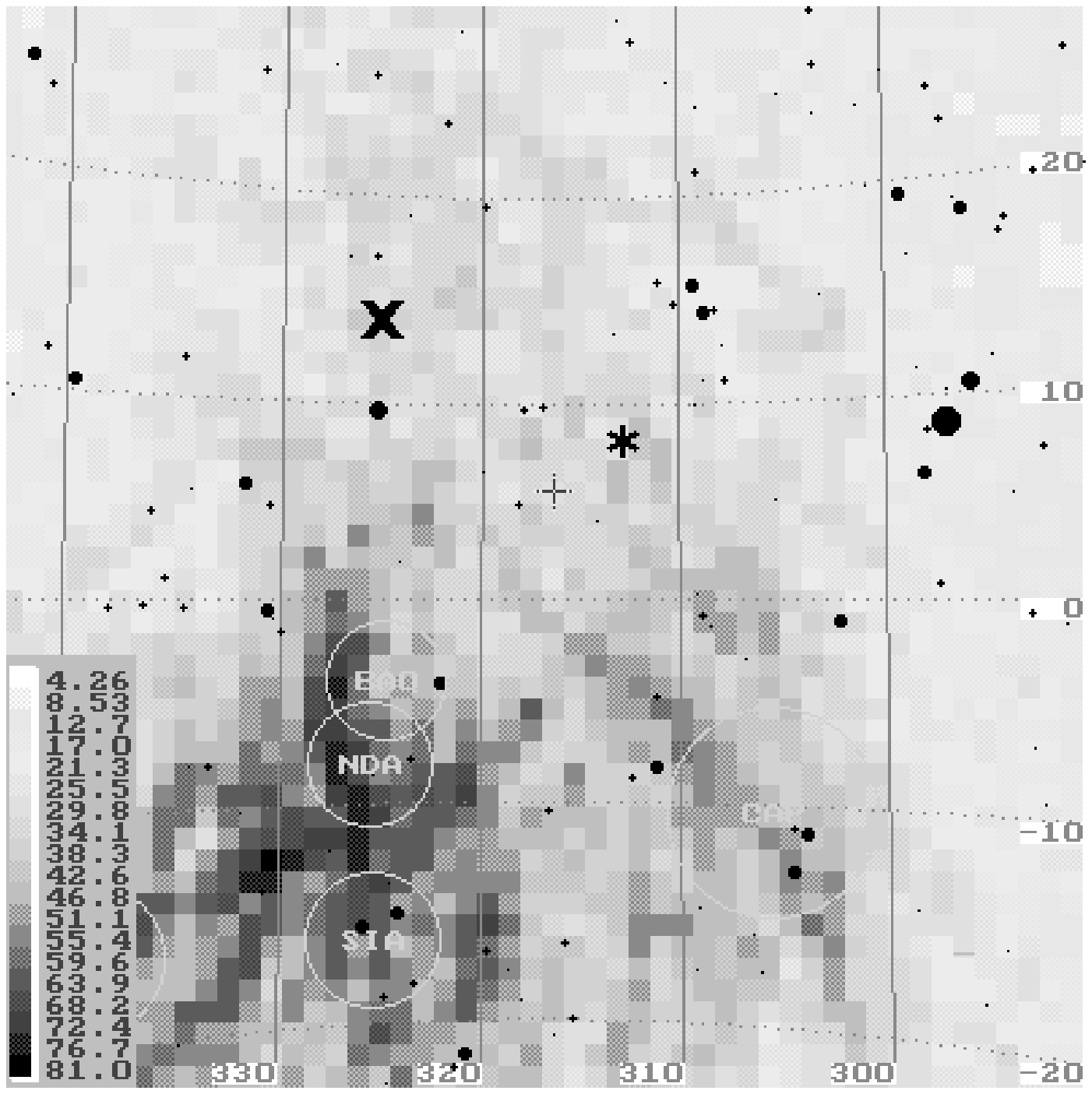}
\includegraphics[scale=.6]{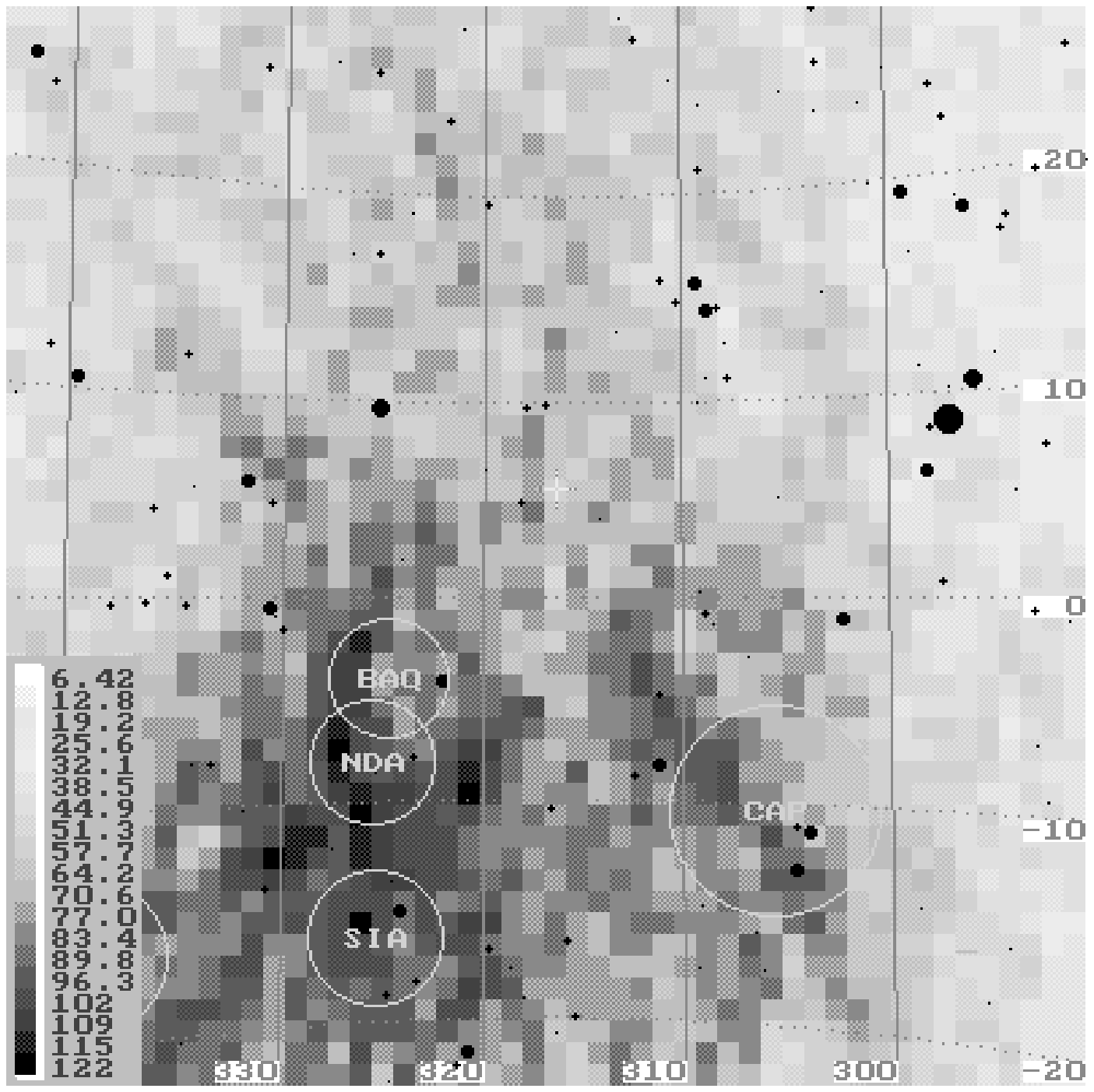}
      \caption{Tracing maps for an artificial sample of 20376 meteors.
All maps are computed for the following parameters:
$\lambda_\odot=125^\circ$, $\Delta\lambda=1.0^\circ$ and $V_\infty=35$
km/s. The maximum distance of the meteor from the radiant is 50 and 
$85^\circ$, respectively from the upper to lower panel.
The big cross marks the position of antihelion source and the asterisk the mean
position of Delpninids' radiant derived from real observations of {\sl CMW}
              }
         \label{FigVibStab}
   \end{figure}

\clearpage

\noindent also detected in the northern part of all figures caused by
the putting a large and diffuse radiant in the zenith.

All these PPR maps strongly suggest that Delphinids indeed exist  
but detecting this shower, due to the vicinity of ecliptic showers and  
the antihelion source, requires a large sample of good quality data.

Following the approach done for the $\alpha$-Cygnids and described in the
previous paragraph, we computed the tracings maps for both real and
simulated samples. The results are shown in Figs. 15 and 16. Because the
top panels, computed for meteors placed within 50 degrees from the
center of the map, suffer less from the meteors radiating from other
showers and sources, we focus our discussion on them.

In the upper panel of Fig. 15 (we do not mark the position of the
Delphinids' radiant for clarity) we detect a clear clump of bright pixels
at the Delphinids' radiant. They correspond to 15-18 intersections
suggesting the presence of the real radiant.

In the tracings map obtained from the simulated sample the strongest
traces, indicating a high number of intersections, come from the real
radiants of the Aquarid complex. We also detect a wide and significantly
less clear tail elongated toward the antihelion source. This tail
reaches to the position of the radiant of the Delphinids shower and at
this location shows about 30-40 intersections. It is less than half of
the number of intersections detected in the area around the radiant of
the Aquarid complex. In the case of the real sample, the trace of the
Delphinids radiant was at the same level as the Aquarids and therefore
we conclude that the tracings maps also indicate that the Delphinids are
a real shower.

\section{Discussion}

We have compared two samples containing meteors observed in July in
years 1996-1999. In the real sample, obtained from real visual
observations made by Polish amateur astronomers, we detected clear
radiants of $\alpha$-Cygnid and Delphinid showers. The question, which
we wanted to answer was whether these radiants can be produced as the
intersections of paths of meteors radiating from the real showers active
in July. Thus we constructed the artificial sample resembling in all
details the real observations and we included all meteor showers except
the $\alpha$-Cygnids and the Delphinids. These radiants, assuming that
they are artificial formations created by intersections of meteors from
real showers, should also be seen in the simulated sample. A comparison
of both databases showed that it is very difficult to produce circular
and clear radiants of the $\alpha$-Cygnids and the Delphinids using the
meteors from an artificial sample. On the other hand such radiants are
easy detected in the real sample. This strongly supports the hypothesis
that the $\alpha$-Cygnids and the Delphinids indeed exist.

Finally, we decided to perform another test. To the artificial sample we
added the meteors from the $\alpha$-Cygnids and the Delphinids. These
showers were described by the parameters ${\rm ZHR}_{\rm max}$, $B$ and
$\lambda_{\odot (max)}$ given by Stelmach \& Olech (2000) and
Wi\'sniewski \& Olech (2001). The numbers of our artificial sample were
then increased by 2035 $\alpha$-Cygnids and 704 Delphinids.


   \begin{figure}[h]
   \centering
\includegraphics[scale=.59]{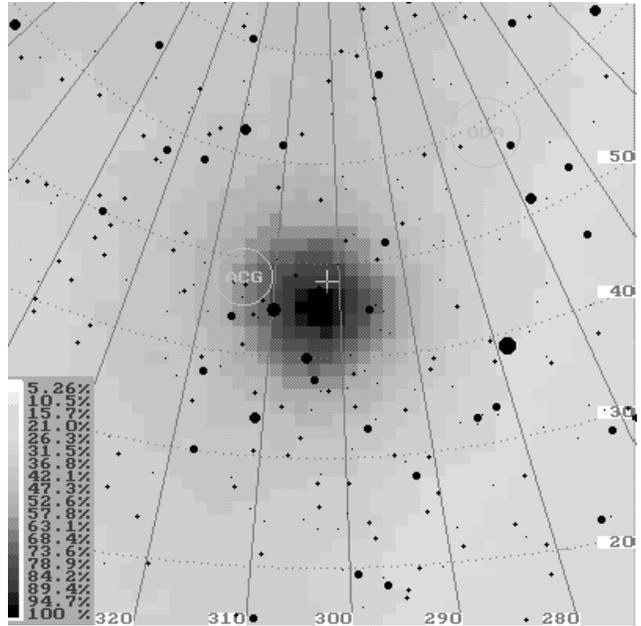}
      \caption{The PPR map for an artificial sample
of 24341 meteors with the $\alpha$-Cygnids and the Delphinids included and
computed for the following parameters:
$\lambda_\odot=115^\circ$, $\Delta\lambda=1.0^\circ$ and $V_\infty=41$
km/s. The maximum distance of the meteor from the radiant is $70^\circ$.
              }
         \label{FigVibStab}
   \end{figure}


   \begin{figure}[h]
   \centering
\includegraphics[scale=.59]{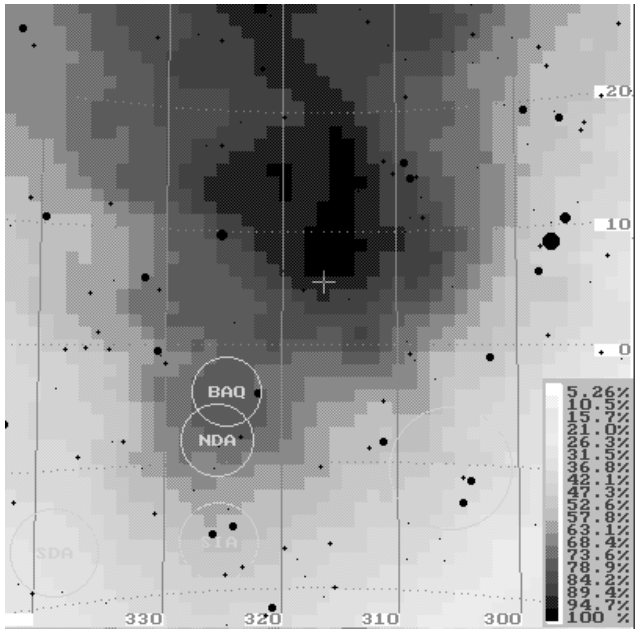}
      \caption{The PPR map for an artificial sample of 23076 meteors
with the $\alpha$-Cygnids and the Delphinids included and
computed for the following parameters:
$\lambda_\odot=125^\circ$, $\Delta\lambda=1.0^\circ$ and $V_\infty=35$
km/s. The maximum distance of the meteor from the radiant is $70^\circ$.
              }
         \label{FigVibStab}
   \end{figure}

We calculated the PPR maps for this new database and they are presented
in Figs. 17 and 18 (for simplicity we decided to compute the maps for
meteors within 70 degrees from the radiant only). Assuming that there  
are no other showers in July than the Delphinids, $\alpha$-Cygnids and
these listed in Table 1 we expect that our artificial sample should
produce the same PPR maps as the real sample (shown in Figs. 9 and 13).

In the case of the $\alpha$-Cygnids we see that the artificial radiant
is more compact than the one obtained from the real sample. However we
should expect that there are few poorly known or even unknown showers
which are present in the real sample and which were not included in the
artificial sample. A good example is the $o$-Draconids shower, which is
not listed in the IMO Working List of Meteor Showers, and as it is
clearly visible from Fig. 9 is detected in our visual data causing a
strong disturbance into the shape of the $\alpha$-Cygnids radiant.

In the case of the Delphinids both Fig. 18 and especially lower panel of
Fig. 13 are similar. The radiant of the Delphinid shower is elongated
toward the $\alpha$-Cygnids radiant. Also the influence of the Aquarid
complex is present in both cases.

The similarity of the PPR maps obtained from the new artificial and real
samples is another argument for the existence of the $\alpha$-Cygnids and
the Delphinids.

Our artificial databases are accessible via Internet and can be
downloaded from the following URL:
http://www.astrouw.edu.pl/$\sim$olech/SIM/.
Detailed information about these databases are included in the README
file.

\begin{acknowledgements}

We are grateful to Chris O'Connor and Dr. Marcin Kiraga for reading and
commenting on this manuscript. This work was supported by KBN grant
number 5~P03D~020~26. AO also acknowledges support from Fundacja na
Rzecz Nauki Polskiej. Additionaly we would like to thank Prof. Marcin
Kubiak from Warsaw University Observatory for allowing us to organize
the {\sl CMW} astronomical camps in Ostrowik where the majority of our
data was collected.

\end{acknowledgements}

\end{document}